\documentclass[a4paper]{article}
\usepackage{amsxtra,amssymb,amsmath,enumerate,graphicx,amsthm,color}
\usepackage[T1]{fontenc}
\usepackage[utf8]{inputenc}
\usepackage{lmodern}
\usepackage[english]{babel}
\usepackage{csquotes}
\usepackage[active]{srcltx}
\usepackage{t1enc}
\usepackage{aurical}
\usepackage{float}
\usepackage{subcaption}
\usepackage{titlesec}
\usepackage{float}
\usepackage{booktabs}
\usepackage{multirow}
\usepackage{accents}
\usepackage{tabularx}
\usepackage[utf8]{inputenc}
\usepackage{apacite}
\usepackage{geometry}
\usepackage{listings} 
\usepackage[authoryear]{natbib}
\usepackage[unicode=true,pdfusetitle,
 bookmarks=true,bookmarksnumbered=true,bookmarksopen=true,bookmarksopenlevel=1,
 breaklinks=true, backref=page, pdfborder={0 0 1},colorlinks=true]
 {hyperref}
\hypersetup{
 linkcolor=blue, citecolor=blue, urlcolor=blue, filecolor=blue,pdfpagelayout=OneColumn, pdfnewwindow=true,pdfstartview=XYZ, plainpages=false}
\usepackage{breakurl}

\providecommand{\conditionname}{Condition}
\providecommand{\lemmaname}{Lemma}
\providecommand{\conditionname}{Condition}
\providecommand{\propositionname}{Proposition}

\theoremstyle{definition}

\theoremstyle{plain}

\theoremstyle{definition}

\theoremstyle{definition}
\newtheorem{defn}{\protect\definitionname}[section]
\theoremstyle{remark}
\newtheorem{rem}{\protect\remarkname}[section]

\usepackage{enumitem}
\setlist{leftmargin=*, topsep=0.5em, parsep=0pt, itemsep=1em, labelindent=0pt, align=left}

\makeatletter

\usepackage{ifpdf} 
\ifpdf 

 \IfFileExists{lmodern.sty}{\usepackage{lmodern}}{}

\fi 

\let\myTOC\tableofcontents
\renewcommand\tableofcontents{%
  \pdfbookmark[1]{\contentsname}{}
  \myTOC
}

\def\LyX{\texorpdfstring{%
  L\kern-.1667em\lower.25em\hbox{Y}\kern-.125emX\@}
  {LyX}}

\@addtoreset{footnote}{section}

\usepackage{dsfont}
\usepackage{bbm}

\usepackage{xcolor}
\definecolor{blue}{HTML}{1F77B4}
\definecolor{orange}{HTML}{FF7F0E}
\definecolor{green}{HTML}{2CA02C}

\usepackage[algo2e,ruled,vlined]{algorithm2e}
\definecolor{algoColorKeyword}{named}{blue}
\definecolor{algoColorComment}{named}{olive}

\SetKwComment{Comment}{$\triangleright$\ }{}

\renewcommand*{\backrefalt}[4]{%
   \ifcase #1 
    \or 
      (cited on page~#2)%
   \else
      (cited on pages~#2)
    \fi}

\makeatother
\providecommand{\definitionname}{Definition}
\providecommand{\remarkname}{Remark}

\begin{document}

\title{Fast multivariate empirical cumulative distribution
function\\ 
with connection to kernel density estimation}
\author{Nicolas Langren\'e\thanks{CSIRO Data61, RiskLab Australia,  nicolas.langrene@csiro.au},\ \ 
Xavier Warin\thanks{EDF Lab, FiME (Laboratoire de Finance des March\'es de l'\'Energie),
warin@edf.fr}}
\date{May 25, 2020}
\maketitle
\begin{abstract}
This paper revisits the problem of computing empirical cumulative distribution functions (ECDF) efficiently on large, multivariate datasets. Computing an ECDF at one evaluation point requires $\mathcal{O}(N)$ operations on a dataset composed of $N$ data points. Therefore, a direct evaluation of ECDFs at $N$ evaluation points requires a quadratic $\mathcal{O}(N^2)$ operations, which is prohibitive for large-scale problems. Two fast and exact methods are proposed and compared. The first one is based on fast summation in lexicographical order, with a $\mathcal{O}(N{\log}N)$ complexity and requires the evaluation points to lie on a regular grid. The second one is based on the divide-and-conquer principle, with a $\mathcal{O}(N\log(N)^{(d-1){\vee}1})$ complexity and requires the evaluation points to coincide with the input points. The two fast algorithms are described and detailed in the general $d$-dimensional case, and numerical experiments validate their speed and accuracy.
Secondly, the paper establishes a direct connection between cumulative distribution functions and kernel density estimation (KDE) for a large class of kernels. This connection paves the way for fast exact algorithms for multivariate kernel density estimation and kernel regression. Numerical tests with the Laplacian kernel validate the speed and accuracy of the proposed algorithms. A broad range of large-scale multivariate density estimation, cumulative distribution estimation, survival function estimation and regression problems can benefit from the proposed numerical methods.\vspace{1em}

\vspace{1em}

\end{abstract}

\section{Introduction\label{sec:Introduction}}

Let $(x_{1},y_{1}),(x_{2},y_{2}),\ldots,(x_{N},y_{N})$ be a sample
of $N$ input (source) points $x_{i}=(x_{1,i},x_{2,i},\ldots,x_{d,i})\in\mathbb{R}^{d}$
and output points $y_{i}\in\mathbb{R}$. Consider an evaluation (target)
point $z=(z_{1},z_{2},\ldots,z_{d})\in\mathbb{R}^{d}$. We define
a generalized multivariate empirical cumulative distribution function
(ECDF) as follows:
\begin{equation}
F_{N}(z)=F_{N}(z;x,y)\triangleq\frac{1}{N}\sum_{i=1}^{N}y_{i}\mathbbm{1}\{x_{1,i}\leq z_{1},\ldots,x_{d,i}\leq z_{d}\}\,.\label{eq:ECDF}
\end{equation}
In a similar manner, we define a generalized multivariate empirical
survival function (ESF) (a.k.a. complementary cumulative distribution
function) as follows:
\begin{equation}
\bar{F}_{N}(z)=\bar{F}_{N}(z;x,y)\triangleq\frac{1}{N}\sum_{i=1}^{N}y_{i}\mathbbm{1}\{x_{1,i}>z_{1},\ldots,x_{d,i}>z_{d}\}\,.\label{eq:ESF}
\end{equation}
The particular case $y\equiv1$ corresponds to the classical joint
empirical distribution function $F_{N}(z)=\frac{1}{N}\sum_{i=1}^{N}\mathbbm{1}\{x_{1,i}\leq z_{1},\ldots,x_{d,i}\leq z_{d}\}$.

More generally, define the following multivariate ECDF:
\begin{align}
F_{N}(z,\delta) & =F_{N}(z,\delta;x,y)\triangleq\frac{1}{N}\sum_{i=1}^{N}y_{i}\mathbbm{1}\{x_{1,i}\leq_{\delta_{1}}z_{1},\ldots,x_{d,i}\leq_{\delta_{d}}z_{d}\}\label{eq:ECDFdelta}
\end{align}
where $\delta=\left\{ \delta_{1},\delta_{2},\ldots,\delta_{d}\right\} \in\left\{ -1,1\right\} ^{d}$,
and where the generalized inequality operator $\leq_{c}$ corresponds
to $\leq$ (lower or equal) if $c\geq0$, and to $<$ (strictly lower) if $c<0$. In particular $F_{N}(z)=F_{N}(z,1;x,y)$
and $\bar{F}_{N}(z)=F_{N}(-z,-1;-x,y)$ respectively.

Cumulative distribution functions and their empirical counterparts
are a cornerstone of statistical theory. In particular, classical
statistical tests of equality of probability distributions such as
the Kolmogorov-Smirnov, Cram\'er-von Mises and Anderson-Darling tests
are based upon empirical distribution functions \citep{Green1976}.

The multivariate versions of these tests are methodologically and
computationally more involved (\citealt{Justel1997}, \citealt{Chiu2009})
due to the greater complexity of multivariate ECDFs \eqref{eq:ECDF}
compared to their univariate counterpart.

A copula is a particular case of multivariate cumulative distribution
function with uniform marginals \citep{Durante2010}. Empirical copulas
appear in the computation of multivariate measures of association
(generalizing the bivariate Spearman rho, \citealt{Schmid2007}, \citealt{Schmid2010}).

The focus of this article is on the numerical computation of generalized
multivariate empirical cumulative distribution functions as defined
in equation \eqref{eq:ECDFdelta}. As the computation of the ECDF \eqref{eq:ECDFdelta}
at one evaluation point $z$ requires $\mathcal{O}(N)$ operations,
a direct implementation of equation \eqref{eq:ECDFdelta} on a set of
$M$ evaluation points requires $\mathcal{O}(M\times N)$ operations.
In particular, when the evaluation points coincide with the input
points $x_{1},x_{2},\ldots,x_{N}$, a direct evaluation requires a
quadratic $\mathcal{O}(N^{2})$ operations. 

The main contribution of this article is to propose an exact $\mathcal{O}(N\log N)$
algorithm to perform this task, based on independent data sorting
in each dimension, combined with a fast lexicographical-sweep summation algorithm
(subsection \ref{subsec:Fast-lexicographical-sweep}). Should the
input data be already sorted, the computational complexity is reduced
to an optimal $\mathcal{O}(N)$. This new algorithm is compared with
the state-of-the-art for fast multivariate ECDF computation, namely
the fast divide-and-conquer recursion of \citet{Bentley1980} with
$\mathcal{O}(N\log(N)^{(d-1){\vee}1})$ computational complexity (subsection
\ref{subsec:Fast-divide-and-conquer}).

The second main contribution of this article is to establish that
a large class of kernel density estimators can be decomposed into
a sum of ECDFs (subsection \ref{subsec:CDF_decomposition_KDE}), which
yields an exact $\mathcal{O}(N\log N)$ kernel density estimation
approach in the lines of \citet{Langrene2019}, as well as a novel
$\mathcal{O}(N\log(N)^{(d-1){\vee}1})$ kernel density estimation algorithm based
on the divide-and-conquer approach of \citet{Bentley1980}. The table below summarizes the contributions of this paper.

\begin{center}
\begin{tabular}{lll}
\hline
Contributions & multivariate CDF & multivariate KDE\tabularnewline
\hline 
fast summation & this paper & \citet{Langrene2019}\tabularnewline
divide-and-conquer & \citet{Bentley1980} & this paper\tabularnewline
\hline 
\end{tabular}
\par\end{center}

The class
of compatible kernels contain popular kernels such as the uniform,
Epanechnikov and Laplacian kernels (subsection \ref{subsec:compatible-kernels}).
It also contains a large class of polynomial-exponential kernels which can
be used to uniformly approximate any incompatible kernel such as the
Gaussian kernel to arbitrary precision (subsection \ref{subsec:new-laplacian}).

The numerical tests reported in Section \ref{sec:numerics} illustrate
the speed and accuracy of the proposed numerical methods. In practice,
the fast summation algorithm requires the evaluation points
to lie on a rectilinear grid, while the divide-and-conquer algorithm
requires the evaluation points to be the same as the input points.
These constraints mean that depending on the chosen algorithm and
the set of evaluation points, an additional interpolation of the results
might be necessary, the impact of which on accuracy can be deemed acceptable
(Figures \ref{fig:errorFastCDF} and \ref{fig:errorFastKDE} in Section \ref{sec:numerics}). 

The contributions of this article can benefit any numerical procedure
requiring a nonparametric estimation of univariate or multivariate
cumulative density functions, survival functions or probability density functions. In
particular, statistical tests of equality of probability distributions
\citep{Green1976}, nonparametric empirical copula estimation \citep{Choros2010}, kernel density estimation and kernel regression all benefit from the proposed fast computation
of ECDFs.

\section{Fast computation of multivariate cumulative distribution\label{sec:FastCDF}}

This section presents two fast algorithms to compute the generalized
empirical distributions (\ref{eq:ECDF}-\ref{eq:ESF}-\ref{eq:ECDFdelta}).

The first one is based on the fast sum updating idea (\citealt{Chen2006},
\citealt{Langrene2019}). It requires a rectilinear evaluation grid,
and its computational complexity is $O(N\log N)$, or $O(N)$ in the
case of a uniform grid. It is described in subsection \ref{subsec:Fast-lexicographical-sweep}. To the authors' knowledge, it is the first time this computational technique is used to compute multivariate ECDFs.

The second one is based on the divide-and-conquer principle (\citealt{Bentley1980}, \citealt{bouchard2012monte},
\citealt{Lee2018}). It requires the evaluation points to be equal
to the input points, and its computational complexity is $O(N\log(N)^{(d-1){\vee}1})$ where $d$ is the dimension of the multivariate input data.
It is described in subsection \ref{subsec:Fast-divide-and-conquer}. 

Another fast ECDF algorithm proposed in the literature can be found in \citet{Perisic2005}; however, this algorithm has been specifically designed for the bivariate case and cannot be extended to higher dimensional ECDFs.

\subsection{Fast sum updating in lexicographical order\label{subsec:Fast-lexicographical-sweep}}

Let $z_{j}=(z_{1,j},z_{2,j},\ldots,z_{d,j})\in\mathbb{R}^{d}$, $j\in\{1,2,\ldots,M\}$,
be a set of $M$ evaluation (target) points.

We require this evaluation grid to be rectilinear, i.e., the $M$
evaluation points $z_{1},z_{2},\ldots,z_{M}$ lie on a regular grid
with possibly non-uniform mesh, of dimension $M_{1}\times M_{2}\times\ldots\times M_{d}=M$:
\[
\mathbf{z}=\left\{ (z_{1,j_{1}},z_{2,j_{2}},\ldots,z_{d,j_{d}})\in\mathbb{R}^{d},\,j_{k}\in\{1,2,\ldots,M_{k}\},\,k\in\{1,2,\ldots,d\}\right\} 
\]
For convenience, we extend the definition of the grid with the notational conventions $z_{k,0}\triangleq-\infty$ and $z_{k,M_{k}+1}\triangleq\infty$.

In each dimension $k\in\{1,2,\ldots,d\}$, the vector $(z_{k,1},z_{k,2},\ldots,z_{k,M_{k}})\in\mathbb{R}^{M_{k}}$
is assumed to be sorted in increasing order:
\[
z_{k,1}<z_{k,2}<\ldots<z_{k,M_{k}},\,k\in\{1,2,\ldots,d\}
\]
We partition the input data $\mathbf{x}$ along this evaluation grid
$\mathbf{z}$. For each evaluation grid index $(j_{1},j_{2},\ldots,j_{d})\in\{1,2,\ldots,M_{1}+1\}\times\ldots\times\{1,2,\ldots,M_{d}+1\}$
we define the following local sum
\begin{equation}
s_{j_{1},j_{2},\ldots,j_{d}}:=\frac{1}{N}\sum_{i=1}^{N}y_{i}\mathbbm{1}\{z_{1,j_{1}-1}<x_{1,i}\leq z_{1,j_{1}},\ldots,z_{d,j_{d}-1}<x_{d,i}\leq z_{d,j_{d}}\}\label{eq:partial_sums}
\end{equation}
Together, the sums \eqref{eq:partial_sums}
form a generalized multivariate histogram (classical histogram
in the case $y\equiv1$). For completeness, the computation of the local sums \eqref{eq:partial_sums}
is detailed in Appendix \ref{sec:local-sums}.

In particular, using equation \eqref{eq:ECDF}, the following key
equality holds:
\begin{equation}
F_{N}(z)=\sum_{l_{1}=1}^{j_{1}}\!\sum_{l_{2}=1}^{j_{2}}\cdots\sum_{l_{d}=1}^{j_{d}}s_{l_{1},l_{2},\ldots,l_{d}}\label{eq:ECDF_decomposition}
\end{equation}
for any evaluation point $z=(z_{1,j_{1}},z_{2,j_{2}},\ldots,z_{d,j_{d}})\in\mathbf{z}$.
We propose a simple fast summation algorithm, Algorithm \ref{algo:fast_sum_CDF}, to compute the ECDFs $F_N(z)$ for every $z\in\mathbf{z}$ in lexicographical order based on the local sum decomposition \eqref{eq:ECDF_decomposition}. One can easily verify that the number of operations is proportional to $M_{1}\times M_{2}\times\ldots\times M_{d}=M$. As Appendix \ref{sec:local-sums} shows that the computation of the local sums \eqref{eq:partial_sums} costs $\mathcal{O}(N{\log}N)$ operations (or only $\mathcal{O}(N)$ if the grid is uniform or the data already sorted), the overall computational complexity of Algorithm \ref{algo:fast_sum_CDF} is $\mathcal{O}(M+N{\log}N)$, or $\mathcal{O}(N{\log}N)$ when $M{\approx}N$ (respectively $\mathcal{O}(M+N)$ and $\mathcal{O}(N)$ when the grid is uniform or the data already sorted).\\

\begin{rem}\label{rem:1N_scaling}
One can alternatively define the local sums \eqref{eq:partial_sums}
without the $1/N$ scaling factor, and apply the division by $N$
to the output of Algorithm \ref{algo:fast_sum_CDF} (equation \eqref{eq:ECDF_decomposition}). This modification
ensures Algorithm \ref{algo:fast_sum_CDF} does not generate any float
rounding error in the case when the $y_{i}$ take integer values,
which includes the classical CDF case $y\equiv1$.
\end{rem}

\begin{algorithm2e}[H]
\DontPrintSemicolon 
\SetAlgoLined 

\vspace{1mm}

\KwIn{precomputed sums $s_{l_{1},l_{2},\ldots,l_{d}}$}

\vspace{0.5mm}

$\mathcal{S}_{1,l_{2},l_{3},\ldots,l_{d}}=0$

\For{$\textcolor{blue}{\mathbf{(}}$  $j_{1}=1,...,M_{1}+1$ $\textcolor{blue}{\mathbf{)}}$}{
\vspace{0.5mm}

$\mathcal{S}_{1,l_{2},l_{3},\ldots,l_{d}}\,\,+\!=\,s_{j_{1},l_{2},l_{3},\ldots,l_{d}}$,~~${\scriptstyle \forall l_{k}\in\left\{ 1,2,\ldots, M_{k}+1 \right\} }$,~${\scriptstyle k\in\{2,3,\ldots,d\}}$\vspace{0.5mm}

\Comment*[l]{Here $\mathcal{S}_{1,l_{2},l_{3},\ldots,l_{d}}\!=\sum_{l_{1}=1}^{j_{1}}\!s_{l_{1},l_{2},\ldots,l_{d}}$,~${\scriptstyle \forall l_{k}\in\left\{ 1,2,\ldots,M_{k}+1\right\} }$,~${\scriptstyle k\in\{2,3,\ldots,d\}}$}

\vspace{1mm}

$\mathcal{S}_{2,l_{3},\ldots,l_{d}}=0$

\For{$\textcolor{blue}{\mathbf{(}}$ $j_{2}=1,...,M_{2}+1$ $\textcolor{blue}{\mathbf{)}}$}{

\vspace{0.5mm}

$\mathcal{S}_{2,l_{3},\ldots,l_{d}}\,\,+\!=\,\mathcal{S}_{1,j_{2},l_{3},\ldots,l_{d}}$,~~${\scriptstyle \forall l_{k}\in\left\{ 1,2,\ldots,M_{k}+1\right\} }$,~${\scriptstyle k\in\{3,\ldots,d\}}$

\vspace{0.5mm}

\Comment*[l]{Here $\mathcal{S}_{2,l_{3},\ldots,l_{d}}\!=\sum_{l_{1}=1}^{j_{1}}\!\sum_{l_{2}=1}^{j_{2}}\!\!s_{l_{1},l_{2},\ldots,l_{d}}$,~${\scriptstyle \forall l_{k}\in\left\{ 1,\ldots, M_{k}+1\right\} }$,~${\scriptstyle k\in\{3,\ldots d\}}$}

$\vdots$

$\mathcal{S}_{d}=0$

\For{$\textcolor{blue}{\mathbf{(}}$ $j_{d}=1,...,M_{d}+1$ $\textcolor{blue}{\mathbf{)}}$}{

\vspace{0.5mm}

$\mathcal{S}_{d}\,\,+\!=\,\mathcal{S}_{d-1,j_{d}}$ \\

\vspace{0.5mm}

\Comment*[l]{Here $\mathcal{S}_{d}=\sum_{l_{1}=1}^{j_{1}}\!\sum_{l_{2}=1}^{j_{2}}\cdots\sum_{l_{d}=1}^{j_{d}}s_{l_{1},l_{2},\ldots,l_{d}}$}
\vspace{0.5mm}
\Comment*[l]{$=F_{N}(z_{1,j_{1}},z_{2,j_{2}},\ldots,z_{d,j_{d}})=F_{N}(z)$ from equation \eqref{eq:ECDF_decomposition}}

\vspace{1mm}

} 

} 

} 

\KwOut{$F_{N}(z)$ for all $z\in\mathbf{z}$} 

\vspace{1mm}

\caption{Fast joint empirical cumulative distribution function\label{algo:fast_sum_CDF}}

\end{algorithm2e}

\subsection{Fast divide-and-conquer recursion\label{subsec:Fast-divide-and-conquer}}

Consider the case when the evaluation points $z_j$ are equal to the input points $x_i$. The calculation of the ECDFs $\{F_{N}(x_{i})\}_{i=1,N}$ (equation \eqref{eq:ECDF}) corresponds to a domination
problem in dimension $d$. An algorithm based on a recursive divide-and-conquer sequence has first been proposed in \citet{Bentley1980} for this problem.
An adaptation was proposed in \citet{bouchard2012monte} to solve this problem for
the case of the calculation of conditional expectation using Malliavin
weights. The computational complexity was shown to be $O(c(d)N\log(N)^{(d-1){\vee}1})$.
This algorithm has been rediscovered recently in \citet{Lee2018}.
They give an extensive study based on the quicksort algorithm providing an optimized version of the algorithm of \citet{Bentley1980} and \citet{bouchard2012monte}. Then they extend the approach to the mergesort algorithm.\\
In all the aforementioned papers, although the different authors insist that the algorithm can be generalized in any dimension, the algorithm descriptions are restricted to dimension 3 for the sake of clarity and simplicity.
In the sequel we choose to provide the general $d$-dimensional version of this important algorithm, and refer to the aforementioned papers for the general conceptual ideas about the divide-and-conquer approach to this problem.\\
The pseudo-code is organized as follows: Algorithm \ref{algo:divideMainECDF} is the main function call, which triggers the divide-and-conquer recursive algorithm \ref{algo:divideAndConquerECDF} w.r.t. dimension, starting from the last dimension. At each recursive iteration, the merge algorithm \ref{algo:MergeNDECDF} is used in dimensions below the current dimension. The special 2D case is dealt with the call of the 1D merge algorithm \ref{algo:Merge1D}. Further details regarding how the algorithm works:
\begin{itemize}
\item The $n$-dimensional merge algorithm \ref{algo:MergeNDECDF} is defined using two sets of points $\kappa_1$ and $\kappa_2$ such that each point of $\kappa_2$ dominates the points of $\kappa_1$ in the dimension above the current one $I_{dim}$. A divide-and-conquer algorithm is used in the current dimension, splitting $\kappa_1$ (respectively $\kappa_2$) into two sets $\kappa_{1,1}$ and $\kappa_{1,2}$ (respectively  $\kappa_{2,1}$ and $\kappa_{2,2}$) where each point in $\kappa_{1,2}  \cup \kappa_{2,2}$ dominates all points in $\kappa_{1,1}$ and $\kappa_{2,1}$ in the current dimension.
\item The $n$-dimensional merge is called recursively in the current dimension organizing a divide-and-conquer algorithm for the couple of sets where no clear dominance is available ($(\kappa_{1,1},\kappa_{2,1})$, $(\kappa_{12},\kappa_{2,2})$).
\item For the couple of sets where dominance is clear in the current dimension $(\kappa_{1,1},\kappa_{2,2})$, the $n$-dimensional merge algorithm is called in the dimension below. In the case when $I_{dim}=2$, a direct call to the one-dimensional merge algorithm \ref{algo:Merge1D} is performed.
\end{itemize}

Note that in the algorithm given below, we compute the $F_N$ version excluding the current point. Adding the self contribution for all $F_N$ is linear in time. In addition, some tests to check that sets are not empty are omitted for conciseness.\\

\begin{algorithm2e}[H]
\DontPrintSemicolon 
\SetAlgoLined 
\vspace{1mm}
\KwIn{ $x = (x_i,  \dots, x_N)$, $y = (y_i,  \dots, y_N)$,  for all $i=1, \dots, N$  }
\vspace{0.5mm}
Calculate the permutation $\phi^j$ , $j=1,\dots, d$ such that $ x_{j,\phi^j(1)} \le  x_{j,\phi^j(2)} \le \dots \le x_{j,\phi^j(N)} $\\
\vspace{0.5mm}
$F(x_i) =0 $ for $i=1, \dots , N$ \\
\vspace{0.5mm}
$\mathbf{RecurSplittingECDF}(x, y,  \phi,  F, N)$\\
\KwOut{$ F(x_i)$ for all $i \in [1,N]$}
\vspace{1mm}
\caption{\label{algo:divideMainECDF}Calculate ECDF $ F(x_j) = \sum_{i=1}^N y_i \mathbbm{1}\{x_{1,i}<x_{1,j},\ldots,x_{d,i} <x_{d,j}\} , \quad j=1,N $ \vspace{0.5mm}}
\end{algorithm2e}

\begin{algorithm2e}[H]
\DontPrintSemicolon 
\SetAlgoLined 
\vspace{1mm}
\KwIn{ $x$, $y$, $F$,   $\phi^j(i)$ for $i=1, M$, $j=1,d$}
\vspace{0.5mm}
\Comment*[l]{Split sorted data in two sets according to last dimension}
$\kappa_1 = \{ \phi^{d}(i), i =1 , \frac{M}{2} \}$, $\phi_1$ with values in $\kappa_1$ s.t. $ x_{j,\phi_1^j(1)} \le  x_{j,\phi_1^j(2)} \le \dots \le x_{j,\phi_1^j(\frac{M}{2})}$, $j=1,d$\\
\vspace{0.5mm}
$\kappa_2 = \{ \phi^{d}(i), i = \frac{M}{2}+1 , M \}$, $\phi_2$  in $\kappa_2$ s.t. $ x_{j,\phi_2^j(1)} \le  x_{j,\phi_2^j(2)} \le \dots \le x_{j,\phi_2^j(\frac{M}{2})}$,$j=1,d$ \\
\vspace{0.5mm}
$\mathbf{RecurSplittingECDF}(x, y,  \phi_1,  F, M/2)$\\
\vspace{0.5mm}
$\mathbf{RecurSplittingECDF}(x, y,  \phi_2,  F, M/2)$ \\
\eIf{$\textcolor{blue}{\mathbf{(}}$ d > 2 $\textcolor{blue}{\mathbf{)}}$}
{
\Comment*[l]{Recursive merge for dimension above 2}
\vspace{0.5mm}
$\mathbf{MergeNDECDF}( x, \phi_1, \phi_2 , d-1, y,F, M/2, M/2 )$
}
{
\Comment*[l]{ Merge 1D}
$\mathbf{Merge1D}(x, \phi_1^1, \phi_2^1, y, F)$,\\
}
\KwOut{ $F$ updated }
\caption{\label{algo:divideAndConquerECDF} Recursive splitting function $\mathbf{RecurSplittingECDF}$}
\end{algorithm2e}
\clearpage
\begin{algorithm2e}[H]
\DontPrintSemicolon 
\SetAlgoLined 
\vspace{1mm}
\KwIn{ $x$, $y$, $F$,  $\phi_1^j(i)$, for all $i=1,M_1$, $\phi_2^j(i)$, for all $i=1,M_2$ with values in $[1,N]$ for $j=1,I_{dim}$}
\vspace{0.5mm}
$\kappa_1 = \{ \phi_1^{I_{dim}}(i), i = 1 , M_1 \}$ , $\kappa_2= 
\{ \phi_2^{I_{dim}}(i), i = 1 , M_2 \}$\\
\vspace{0.5mm}
\Comment*[l]{ Merge the two sets involved and find median coordinate in dimension $I_{dmin}$: linear cost with the number of particles}
$\kappa =  \kappa_1 \cup  \kappa_2$, $x_{med}$ s.t; $\#\{ x_{j}, j \in \kappa ,  x_{I_{dim},j} \le x_{med}\} = \#\{ x_{j}, j \in \kappa ,  x_{I_{dim},j} > x_{med}\}$\\
\vspace{0.5mm}
$\kappa_{l,1} = \{ i \in \kappa_l , x_{I_{dim},i} \le x_{med}\}$,  $M_{l,1}= \# \kappa_{l,1}$, for $l=1,2$, \\
\vspace{0.5mm}
$\kappa_{l,2} = \{ i \in \kappa_l , x_{I_{dim},i} > x_{med}\}$,  $M_{l,2}= \# \kappa_{l,2}$,  for $l=1,2$ \\
\Comment*[l]{Sort each set for all dimension below or equal to  $I_{dim}$ : linear in time using $\phi_1^j(i)$, $\phi_2^j(i)$}
Create $\phi_{l,m}^j(i)$, $i=1,\dots M_{l,m}$  s.t. $\phi_{l,m}^j(i) \in \kappa_{l,m}$, and $$x_{j,\phi_{l,m}^j(1)} \le  x_{j,\phi_{l,m}^j(2)} \le \dots \le x_{j,\phi_{l,m}^j(M_{l,m})}, \mbox{ for } j \le I_{dim}, \quad  l=1,2, \quad m=1,2$$.
\vspace{0.5mm}
\Comment*[l]{ Merge for set $\kappa_{1,l}$ and $\kappa_{2,l}$ for $l=1,2$ in same dimension $I_{dim}$}
$\mathbf{MergeNDECDF}(x, \phi_{1,l}, \phi_{2,l} ,I_{dim}, y,F,  M_{1,l}, M_{2,l})$, for $l=1,2$\\
\vspace{0.5mm}
\Comment*[l]{ Clear dominance relation between set below in current dimension: merge in dimension below}
\eIf{$\textcolor{blue}{\mathbf{(}}$ $I_{dim}==2$ $\textcolor{blue}{\mathbf{)}}$}
{
\Comment*[l]{ Merge the set of 3D problem directly without recursion} 
$\mathbf{Merge1D}(x, \phi_{1,1}^1, \phi_{2,2}^1, y, F)$\\
}
{
\Comment*[l]{Merge in dimension below}
$\mathbf{mergedNDECDF}(x, \phi_{1,1}, \phi_{2,2} ,I_{dim}-1, y,F,  M_{1,l}, M_{2,l})$,\\
}
\vspace{0.5mm}
\KwOut{ $F$ updated }
\caption{\label{algo:MergeNDECDF}Recursive merge nD  $\mathbf{MergeNDECDF}$ in given dimension $I_{dim}$}
\end{algorithm2e}

\begin{algorithm2e}[H]
\DontPrintSemicolon 
\SetAlgoLined 
\vspace{1mm}
\KwIn{ $x$, $y$, $F$,  $\phi_k$ s.t. $\phi_k(i) \le \phi_k(i+1)$, for all $i=1,M_k-1$, $k=1,2$ }
$S =0$, $j=1$ \\
\vspace{0.5mm}
\For{$\textcolor{blue}{\mathbf{(}}$ $i=1,M_2$ $\textcolor{blue}{\mathbf{)}}$}
{
\While{$\textcolor{blue}{\mathbf{(}}$  $(x_{\phi_2(i),1} \ge x_{\phi_1(j),1})$  and $j \le M_1$ $\textcolor{blue}{\mathbf{)}}$}
{
$S +=  y_{\phi_1(j)}$, $j= j+1$
}
$F(\phi_2(i)) += S$ \\
\If{$\textcolor{blue}{\mathbf{(}}$ $j== M_1+1$ $\textcolor{blue}{\mathbf{)}}$}
{
\For{$\textcolor{blue}{\mathbf{(}}$ $k=i+1, M_2$ $\textcolor{blue}{\mathbf{)}}$}
{
$F(\phi_2(k)) += S$ 
}
$i=M_2+1$
}
}
\KwOut{ $F$ updated }
\vspace{0.5mm}
\caption{\label{algo:Merge1D} Final merge function in  dimension one : $\mathbf{Merge1D}$ ,  between $\{ x_{\phi_1(i)}, i=1,M_1 \}$ and  $\{ x_{\phi_2(i)}, i=1,M_2 \}$ }
\end{algorithm2e}

\section{Fast kernel density estimation\label{sec:Fast-KDE}}

This section establishes an explicit connection between the computation
of empirical cumulative distribution functions and the problem of
empirical density estimation, more specifically with kernel density
estimation (KDE). The main consequence of this connection is that
the fast empirical CDF algorithms introduced in Section \ref{sec:FastCDF}
also provide a fast way to compute multivariate kernel density estimators.

\subsection{CDF decomposition of KDE\label{subsec:CDF_decomposition_KDE}}

Using the notations from Section \ref{sec:Introduction}, the (univariate)
weighted kernel density estimator (aka Parzen-Rosenblatt estimator)
at the evaluation point $z$ is given by:
\begin{equation}
\hat{f}_{\mathrm{KDE}}(z):=\frac{1}{N}\sum_{i=1}^{N}w_{i}K_{h}(x_{i}-z)\label{eq:localdens}
\end{equation}
where $K_{h}(u):=\frac{1}{h}K\!\left(\frac{u}{h}\right)$ with kernel
$K$ and bandwidth $h$. The classical KDE estimator corresponds to
the weights $w_{i}\equiv1$. Allowing general weights brings more
flexibility, and does not affect the analysis of this section. For
example $w_{i}$ can contain the value of a response variable, as
in local kernel regression estimation (\citealt{Nadaraya1964}, \citealt{Watson1964}).
Another possible use of $w_{i}$ concerns repeated values: should
the input sample $(x_{1},x_{2},\ldots,x_{N})$ contain repeated values,
one can w.l.o.g. compute the kernel sum \eqref{eq:localdens} on the
unique values of the input sample, weighted by the time each value
appears in the original sample \citep{Titterington1980}. Finally,
this setting also encompasses kernel quantile estimators (\citealt{Parzen1979},
\citealt{Sheather1990}, \citealt{Franke2009}) and some kernel distribution
function estimators (\citealt{Azzalini1981}, \citealt{Kim2005}).

In the following, we focus on the Laplacian kernel, defined by 
\begin{equation}
K(u)=\frac{1}{2}e^{-|u|}\label{eq:Laplacian_kernel_1D}
\end{equation}
Subsections \ref{subsec:compatible-kernels} and \ref{subsec:new-laplacian}
will discuss other possible kernel choices in detail. Following \eqref{eq:localdens},
the Laplacian kernel density estimator is defined by: 

\begin{equation}
\hat{f}_{\mathrm{KDE}}(z)=\frac{1}{N}\sum_{i=1}^{N}\frac{w_{i}}{2h}e^{-\frac{|x_{i}-z|}{h}}\label{eq:Laplacian_kde_1D}
\end{equation}
This kernel density estimator can be decomposed as follows 


\begin{align}
\hat{f}_{\mathrm{KDE}}(z)= & \frac{1}{N}\sum_{i=1}^{N}\frac{w_{i}}{2h}e^{-\frac{|x_{i}-z|}{h}}\nonumber \\
= & \frac{1}{2hN}\left(\sum_{i=1}^{N}w_{i}e^{\frac{x_{i}-z}{h}}\mathbbm{1}\{x_{i}\leq z\}+\sum_{i=1}^{N}w_{i}e^{\frac{z-x_{i}}{h}}\mathbbm{1}\{x_{i}>z\}\right)\nonumber \\
= & \frac{1}{2hN}\left(e^{-\frac{z}{h}}\sum_{i=1}^{N}w_{i}e^{\frac{x_{i}}{h}}\mathbbm{1}\{x_{i}\leq z\}+e^{\frac{z}{h}}\sum_{i=1}^{N}w_{i}e^{-\frac{x_{i}}{h}}\mathbbm{1}\{x_{i}>z\}\right)\nonumber \\
= & \frac{1}{2h}\left(e^{-\frac{z}{h}}F_{N}(z;x,we^{\frac{x}{h}})+e^{\frac{z}{h}}\bar{F}_{N}(z;x,we^{-\frac{x}{h}})\right)\label{eq:Laplacian_decomposition_1d}
\end{align}
where the empirical CDF $F_{N}$ and the empirical complementary CDF
$\bar{F}_{N}$ are defined by equations \eqref{eq:ECDF} and \eqref{eq:ESF}
respectively.

Crucially, such a CDF decomposition of KDE also holds in the multivariate
setting. The multivariate Laplacian kernel is defined by

\begin{equation}
K_{\hspace{-0.0833em}d}(u)=\frac{1}{2^{d}}e^{-|u|}=\frac{1}{2^{d}}e^{-\sum_{k=1}^{d}|u_{k}|}\label{eq:Laplacian_kernel_nD}
\end{equation}
and the weighted multivariate Laplacian kernel density estimator is
given by

\begin{equation}
\hat{f}_{\mathrm{KDE}}(z)=\frac{1}{2^{d}N\Pi_{k=1}^{d}h_{k}}\sum_{i=1}^{N}w_{i}e^{-\left|\frac{x_{i}-z}{h}\right|}=\frac{1}{2^{d}N\Pi_{k=1}^{d}h_{k}}\sum_{i=1}^{N}w_{i}e^{-\sum_{k=1}^{d}\frac{\left|x_{k,i}-z_{k}\right|}{h_{k}}}\label{eq:Laplacian_kde_nD}
\end{equation}

where $h=(h_{1},h_{2},\ldots,h_{d})\in\mathbb{R}^{d}$ is a multivariate
bandwidth. The general matrix bandwidth case is discussed in Appendix
\ref{sec:General-matrix-bandwidth}.

Using the same approach as equation \eqref{eq:Laplacian_decomposition_1d},
the sum \eqref{eq:Laplacian_kde_nD} can be decomposed as follows:
\begin{align}
\hat{f}_{\mathrm{KDE}}(z)= & \frac{1}{2^{d}N\Pi_{k=1}^{d}h_{k}}\sum_{i=1}^{N}w_{i}\prod_{k=1}^{d}\left(e^{-\frac{z_{k}}{h_{k}}}e^{\frac{x_{k,i}}{h_{k}}}\mathbbm{1}\{x_{k,i}\leq z_{k}\}+e^{\frac{z_{k}}{h_{k}}}e^{-\frac{x_{k,i}}{h_{k}}}\mathbbm{1}\{-x_{k,i}<-z_{k}\}\right)\nonumber \\
= & \frac{1}{2^{d}N\Pi_{k=1}^{d}h_{k}}\sum_{i=1}^{N}w_{i}\sum_{\delta\in\{-1,1\}^{d}}\prod_{k=1}^{d}e^{-\frac{\delta_{k}z_{k}}{h_{k}}}e^{\frac{\delta_{k}x_{k,i}}{h_{k}}}\mathbbm{1}\{\delta_{k}x_{k,i}\leq_{\delta_{k}}\delta_{k}z_{k}\}\nonumber \\
= & \frac{1}{2^{d}\Pi_{k=1}^{d}h_{k}}\sum_{\delta\in\{-1,1\}^{d}}e^{-\sum_{k=1}^{d}\frac{\delta_{k}z_{k}}{h_{k}}}\frac{1}{N}\sum_{i=1}^{N}w_{i}e^{\sum_{k=1}^{d}\frac{\delta_{k}x_{k,i}}{h_{k}}}\mathbbm{1}\{\delta_{1}x_{1,i}\leq_{\delta_{1}}\delta_{1}z_{1},\ldots,\delta_{d}x_{d,i}\leq_{\delta_{d}}\delta_{d}z_{d}\}\nonumber \\
= & \frac{1}{2^{d}\Pi_{k=1}^{d}h_{k}}\sum_{\delta\in\{-1,1\}^{d}}e^{-\sum_{k=1}^{d}\frac{\delta_{k}z_{k}}{h_{k}}}F_{N}(\delta z,\delta;\delta x,y)\label{eq:Laplacian_decomposition_nD}
\end{align}
with $y_{i}=y_{i}(\delta):=w_{i}e^{\sum_{k=1}^{d}\frac{\delta_{k}x_{k,i}}{h_{k}}}$,
where we used the definition of the generalized empirical CDF $F_{N}(z,\delta;x,y)$
(equation \eqref{eq:ECDFdelta}) and its generalized inequality operator
$\leq_{c}$.

Equation \eqref{eq:Laplacian_decomposition_nD} shows that the computation
of the multivariate Laplacian kernel density estimator \eqref{eq:kde_nD}
can be decomposed into the computation of $2^{d}$ generalized empirical
CDF \eqref{eq:ECDFdelta}, which can be computed efficiently using
the algorithms described in Section \ref{sec:FastCDF}.

\subsection{Compatible kernels\label{subsec:compatible-kernels}}

In the previous subsection, we used the Laplacian kernel \eqref{eq:Laplacian_kernel_1D}-\eqref{eq:Laplacian_kernel_nD}
to illustrate the concept of CDF decomposition of KDE. Such a decomposition
is not restricted to the Laplacian kernel; actually, a large class
of kernels (though not all kernels) is compatible with such a decomposition.
Let us start with the simplest one, namely the uniform kernel
\begin{equation}
K(u)=\frac{1}{2}\mathbbm{1}{\scriptstyle \{\left|u\right|\leq1\}}\,.\label{eq:uniform_kernel_1D}
\end{equation}
The weighted uniform kernel density estimator is given by
\begin{equation}
\hat{f}_{\mathrm{KDE}}(z)=\frac{1}{N}\sum_{i=1}^{N}\frac{w_{i}}{2h}\mathbbm{1}{\scriptstyle \left\{ \left|\frac{x_{i}-z}{h}\right|\leq1\right\} }\label{eq:uniform_kde_1D}
\end{equation}
and can be decomposed as follows:
\begin{align}
\hat{f}_{\mathrm{KDE}}(z)= & \frac{1}{2hN}\left(\sum_{i=1}^{N}w_{i}\mathbbm{1}{\scriptstyle \left\{ x_{i}\leq z+h\right\} }-\sum_{i=1}^{N}w_{i}\mathbbm{1}{\scriptstyle \left\{ x_{i}<z-h\right\} }\right)\nonumber \\
= & \frac{F_{N}(z+h,1;x,w)-F_{N}(z-h,-1;x,w)}{2h}\,.\label{eq:uniform_decomposition_1D}
\end{align}
The multivariate uniform kernel density estimator is given by
\begin{equation}
K_{\hspace{-0.0833em}d}(u)=\frac{1}{2^{d}}\mathbbm{1}{\scriptstyle \{\Vert u\Vert_{\infty}\leq1\}}\label{eq:uniform_kernel_nD}
\end{equation}
and its corresponding weighted multivariate kernel density estimator
\begin{equation}
\hat{f}_{\mathrm{KDE}}(z)=\frac{1}{2^{d}N\Pi_{k=1}^{d}h_{k}}\sum_{i=1}^{N}w_{i}\mathbbm{1}{\scriptstyle \left\{ \left\Vert\frac{x_{i}-z}{h}\right\Vert_{\infty}\leq1\right\} }=\frac{1}{2^{d}N\Pi_{k=1}^{d}h_{k}}\sum_{i=1}^{N}w_{i}\prod_{k=1}^{d}\mathbbm{1}{\scriptstyle \left\{ \left|\frac{x_{k,i}-z_{k}}{h_{k}}\right|\leq1\right\} }\label{eq:uniform_kde_nD}
\end{equation}
can be decomposed as follows:
\begin{align}
\hat{f}_{\mathrm{KDE}}(z)= & \frac{1}{2^{d}N\Pi_{k=1}^{d}h_{k}}\sum_{i=1}^{N}w_{i}\prod_{k=1}^{d}\left(\mathbbm{1}{\scriptstyle \left\{ x_{k,i}\leq z_{k}+h_{k}\right\} }-\mathbbm{1}{\scriptstyle \left\{ x_{k,i}<z_{k}-h_{k}\right\} }\right)\nonumber \\
= & \frac{1}{2^{d}N\Pi_{k=1}^{d}h_{k}}\sum_{i=1}^{N}w_{i}\sum_{\delta\in\{-1,1\}^{d}}\prod_{k=1}^{d}\delta_{k}\mathbbm{1}\{x_{k,i}\leq_{\delta_{k}}z_{k}+\delta_{k}h\}\nonumber \\
= & \frac{1}{2^{d}\Pi_{k=1}^{d}h_{k}}\sum_{\delta\in\{-1,1\}^{d}}\left(\prod_{k=1}^{d}\delta_{k}\right)F_{N}(z+\delta h,\delta;x,w)\,.\label{eq:uniform_decomposition_nD}
\end{align}
The uniform kernel is the simplest example of the large compatible
class of kernels called symmetric beta kernels (\citealt{Marron1988},
\citealt{Duong2015}), defined in the univariate case by:
\begin{equation}
K(u)=\frac{(1-u^{2})^{\alpha}}{2^{2\alpha+1}B(\alpha+1,\alpha+1)}\mathbbm{1}\{\left|u\right|\leq1\}\label{eq:beta_kernels}
\end{equation}
where we used the Beta function $B(x,y)=\frac{\Gamma(x)\Gamma(y)}{\Gamma(x+y)}$.
This class of kernels includes the uniform ($\alpha=0$), Epanechnikov
($\alpha=1$), biweight ($\alpha=2$) and triweight ($\alpha=3$)
as particular cases. The fast sum updating decompositions in \citet{Gasser1989}
and \citet{Seifert1994} (univariate case) and \citet{Langrene2019}
(multivariate case) can be recognised as CDF decompositions and show
that the class \eqref{eq:beta_kernels} in particular is compatible
with CDF decomposition. While equivalent to fast sum updating decomposition,
one can argue that kernel sum decomposition in terms of CDFs makes
the approach clearer and easier to understand, especially in the multivariate
setting (see equations \eqref{eq:Laplacian_decomposition_nD} and
\eqref{eq:uniform_decomposition_nD}).

In view of this discussion, we can infer from \citet{Langrene2019}
that other kernels such as the tricube kernel $K(u)=\frac{70}{81}(1-\left|u\right|^{3})^{3}\mathbbm{1}\{\left|u\right|\leq1\}$
and the cosine kernel $K(u)=\frac{\pi}{4}\cos\left(\frac{\pi}{2}u\right)\mathbbm{1}\{\left|u\right|\leq1\}$
admit a CDF decomposition of KDE. Kernels based around the Laplacian
kernel, such as the Silverman kernel $K(u)=\frac{1}{2}\exp\left(-\frac{\left|u\right|}{\sqrt{2}}\right)\sin\left(\frac{\left|u\right|}{\sqrt{2}}+\frac{\pi}{4}\right)$
are also compatible, and one can build upon compatible kernels to
create new ones, as shown in subsections \ref{subsec:new-laplacian}
and \ref{subsec:high_order_kernels}.

\subsection{New compatible infinite-support kernels\label{subsec:new-laplacian}}

Unfortunately, some kernels are simply incompatible with CDF decomposition.
They are such that the term $K(\frac{x-z}{h})$ cannot be decomposed
into terms depending on $x$ only and terms depending on $z$ only.
Most incompatible kernels have unbounded support, such as the logistic
kernel $K(u)=\frac{1}{e^{u}+2+e^{-u}}$ , the Cauchy kernel $K(u)=\frac{1}{\pi(1+u^{2})}$,
the Fej\'er-de la Vall\'ee Poussin kernel $K(u)=\frac{1}{\pi}\frac{\sin^{2}(u)}{u^{2}}$,
and most importantly the popular Gaussian kernel $K(u)=\frac{1}{\sqrt{2\pi}}e^{-\frac{u^{2}}{2}}$. 

As the finite-support Epanechnikov kernel $K(u)=\frac{3}{4}\left(1-u^{2}\right)\mathbbm{1}\{\left|u\right|\leq1\}$
is known to be optimal in terms of asymptotic mean integrated squared
error (AMISE, \citealt{Epanechnikov1969}), one can wonder whether
such limitation is actually problematic in practice. However, infinite-support
kernels are not devoid of merit for multiple reasons. For example,
more robust non-asymptotic Fourier-based kernel selection criteria
rule out the Epanechnikov kernel (\citealt{Cline1988}, \citealt{Tsybakov2009})
and recommend infinite-support kernels of Fej\'er type, in particular
the Fej\'er-de la Vall\'ee Poussin kernel (\citealt{Stepanova2013},
\citealt{Kosta2015}). Moreover, infinite-support kernel have been
recommended for consistent likelihood cross-validation (\citealt{Brewer2000},
\citealt{Zhang2006}, \citealt{Hofmeyr2020}), and for tail probability
estimation \citep{Lall1993}. Finally, kernels with unbounded support
produce smooth prediction functions, which is a desirable feature
for density visualization \citep{Berthold2010}.

As pointed out in \citet{Hofmeyr2020}, all known infinite-support
kernels compatible with fast recursions are based around the Laplacian
kernel \eqref{eq:Laplacian_kernel_1D}, which is why subsection \ref{subsec:CDF_decomposition_KDE}
focused on this important kernel. In the multivariate case, infinite-support
kernels are more straightforward to decompose into CDFs than finite-support
kernels. Indeed the decomposition of multivariate Beta kernels in
\citealt{Langrene2019} requires the support of the kernel to be a
hyperrectangle, which holds for product kernels but not for radially
symmetric kernels. By contrast, equation \eqref{eq:Laplacian_decomposition_nD}
shows that obviously no such limitation exists for the Laplacian kernel.

In this subsection, we introduce an important class of kernels which
is compatible with fast recursion and can be used to approximate all
the incompatible kernels mentioned so far. It is defined by
\begin{equation}
K(u)=\frac{\gamma_{p}}{h}\left(\sum_{k=0}^{p}\beta_{k,p}\frac{\left|u\right|^{k}}{h^{k}}\right)e^{-\alpha_{p}\frac{\left|u\right|}{h}}\label{eq:polyexp_kernel}
\end{equation}
with parameters $\alpha_{p}>0$, $\beta_{k,p}$, $k=0,1,\ldots,p$,
and scaling parameter $\gamma_{p}>0$ defined such that the kernel
\eqref{eq:polyexp_kernel} integrates to one:
\begin{equation}
\gamma_{p}\triangleq\frac{1}{2\sum_{k=0}^{p}\beta_{k,p}\frac{k!}{\alpha_{p}^{k+1}}}\,.\label{eq:scaling}
\end{equation}
The bandwidth parameter $h>0$ does not affect the integral of the
kernel ($\int_{-\infty}^{\infty}K(u)du=\frac{1}{h}\int_{-\infty}^{\infty}K\left(\frac{u}{h}\right)du$).
The class of kernels \eqref{eq:polyexp_kernel} contains the Sargan
kernels (\citealt{Goldfeld1981}; or double Gamma kernel sums, \citealt{Nguyen2009}),
and is obtained by multiplying the Laplacian kernel by a polynomial
term in $\left|u\right|$. Such a distribution occurs when averaging
$p+1$ i.i.d. Laplace distributions (\citealt{Craig1932}, \citealt{Weida1935},
\citealt{Kotz2001}).

As pointed out in \citet{Kafaei1985}, the theoretical foundation
for considering kernels of the type \eqref{eq:polyexp_kernel} is
the generalization of the Stone-Weierstrass theorem in \citet[Section 11]{Stone1962}
which states that any continous function can be uniformly approximated
by functions of the form \eqref{eq:polyexp_kernel} (without the scaling
constant). In particular, any continuous density/kernel function can
be uniformly approximated by \eqref{eq:polyexp_kernel} to arbitrary
precision for sufficiently large $p$. This includes all the kernels
incompatible with fast recursion such as the Gaussian kernel.

Indeed, the sub-class of Mat\'ern kernels (\citealt{Matern1960},
\citealt{Matern1986}) defined by 
\begin{align}
K_{\mathrm{Mat\'ern}}(u) & \triangleq\frac{\gamma_{p}}{h}\left(\sum_{k=0}^{p}\frac{p!}{k!(p-k)!}\frac{(2p-k)!}{(2p)!}\left(2\sqrt{2p+1}\right)^{k}\frac{\left|u\right|^{k}}{h^{k}}\right)e^{-\sqrt{2p+1}\frac{\left|u\right|}{h}}\label{eq:Matern_kernel}\\
\gamma_{p} & \triangleq\frac{\sqrt{2p+1}}{2\sum_{k=0}^{p}\frac{p!}{(p-k)!}\frac{(2p-k)!}{(2p)!}2^{k}}\label{eq:Matern_constant}
\end{align}
is known to converge to the Gaussian kernel for large $p$.
\begin{equation}
K_{\mathrm{Mat\'ern}}(u)\underset{p\rightarrow\infty}{\rightarrow}\frac{1}{h\sqrt{2\pi}}e^{-\frac{\left|u\right|^{2}}{2h^{2}}}\label{eq:Matern_to_Gaussian}
\end{equation}
(in particular the scaling constant $\gamma_{p}$ defined in equation
\eqref{eq:Matern_constant} converges to $1/\sqrt{2\pi}$ when $p\rightarrow\infty$).
The case $p=0$ corresponds to the Laplacian kernel \eqref{eq:Laplacian_kernel_1D},
and the Mat\'ern kernels with $p=1$ and $p=2$ are given explicitly
by

\begin{align}
K_{\mathrm{Mat\'ern}\hspace{0.0833em}3/2}(u) & =\frac{\sqrt{3}}{4h}\left(1+\sqrt{3}\frac{\left|u\right|}{h}\right)e^{-\sqrt{3}\frac{\left|u\right|}{h}}\label{eq:Matern32}\\
K_{\mathrm{Mat\'ern}\hspace{0.0833em}5/2}(u) & =\frac{3\sqrt{5}}{16h}\left(1+\sqrt{5}\frac{\left|u\right|}{h}+\frac{5}{3}\frac{\left|u\right|^{2}}{h^{2}}\right)e^{-\sqrt{5}\frac{\left|u\right|}{h}}\label{eq:Matern52}
\end{align}
They are known in the literature as the Mat\'ern-3/2 kernel \eqref{eq:Matern32}
and Mat\'ern-5/2 kernel \eqref{eq:Matern52} respectively, owing
to the classical parameterization $\nu=p+\frac{1}{2}$.

\begin{figure}[H]
\begin{centering}
\includegraphics[width=0.73\paperwidth]{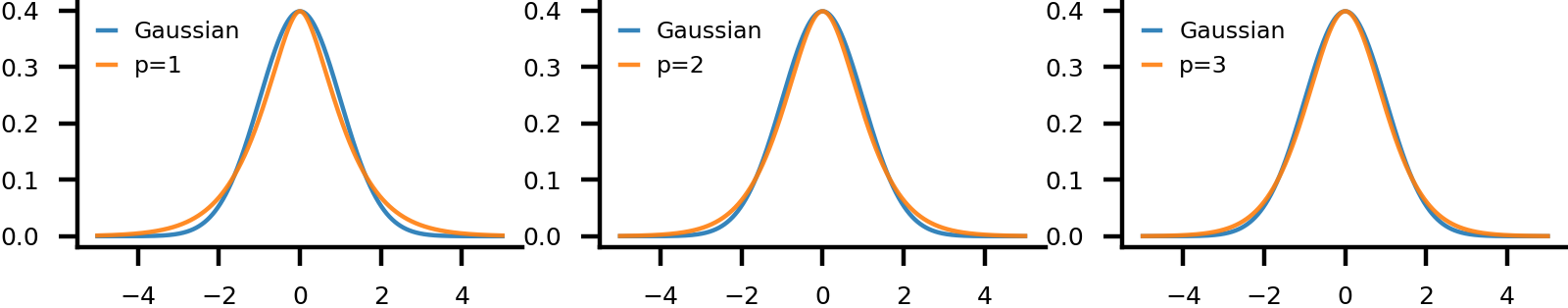}
\par\end{centering}
\caption{Comparison to Gaussian kernel\label{fig:Gaussian_comparison}}
\end{figure}

For the definition of these kernels, the bandwidth parameter $h>0$
can be chosen in different ways: one can fix it to $1$ for simplicity,
to the value defining the canonical shape of the kernel ($h=\left(\int_{\mathbb{R}}K^{2}(u)du\right)^{1/5}/\left(\int_{\mathbb{R}}u^{2}K(u)du\right)^{2/5}$,
\citealt{Marron1988}), or in such a way as to ease the visual comparison
of the kernel shape to some other kernels. As Mat\'ern kernels approximate
Gaussian kernels (equation \eqref{eq:Matern_to_Gaussian}), one can
choose to set $h$ such that $K(0)=1/\sqrt{2\pi}$, namely $h=\sqrt{2\pi}/\gamma_{p}$,
as shown on Figure \ref{fig:Gaussian_comparison}.

In \citet{Goldfeld1981}, the motivation to investigate the class
of distributions \eqref{eq:polyexp_kernel} was to approximate the
Gaussian distribution by a more tractable distribution with explicit
integrals (see also \citealt{Missiakoulis1983}, \citet{Kafaei1985},
\citealt{Tse1987} and \citealt{Hadri1996} for specific kernel suggestions
within the class \eqref{eq:polyexp_kernel}). Figure \ref{fig:Gaussian_comparison}
suggests that computationally-attractive low-order Mat\'ern kernels
such as \eqref{eq:Matern52} or even \eqref{eq:Matern32} might suffice
to approximate the shape of a Gaussian kernel. In the context of kernel
density estimation, the fact that such kernels are compatible with
fast recursions and CDF decompositions make them even more attractive
than Gaussian kernels.

\subsection{New compatible higher-order kernels\label{subsec:high_order_kernels}}

Finally, another interesting set of kernels is the class of higher-order
kernels. 
\begin{defn}
(see for example \citealt{Silverman1986}). A kernel $K$ is said
to be of order $p$ if and only if
\[
\int u^{j}K(u)du=\begin{cases}
1 & \mathrm{if}\,j=0\\
0 & \mathrm{if}\,1\leq j\leq p-1\\
c_{k}\neq0,\left|c_{k}\right|<\infty & \mathrm{if}\,j=p
\end{cases}
\]
\end{defn}
The order $p$ of a kernel is even when $K$ is chosen symmetric.
The kernel order has a direct connection to the best AMISE, namely
$\mathcal{O}\left(N^{-\frac{p}{2p+1}}\right)$, of the KDE estimator
(\citealt{Gasser1985}, \citealt{Silverman1986}). This suggests that
high-order kernels should asymptotically perform better (though see
\citet{Silverman1986} and \citet{Marron1992} on the usefulness of
such kernels on moderate sample sizes).

It is known that any kernel defined as a symmetric probability density
function with finite variance is necessarily of order 2 (\citealt{Schucany1989},
\citealt{Jones1993}). One consequence is that kernels of order $p>2$
necessarily take negative values in places.

Examples of fourth-order kernels include $K(u)=\frac{9}{8}\left(1-\frac{5}{3}u^{2}\right)\mathbbm{1}\{u\leq\left|1\right|\}$
\citep{Bartlett1963}, and $K(u)=\frac{15}{32}\left(3-10u^{2}+7u^{4}\right)\mathbbm{1}\{\left|u\right|\leq 1\}$
\citep{Gasser1985} which, being polynomial kernels, are compatible
with fast recursion (see subsection \ref{subsec:compatible-kernels}).
More generally, there exists various ways to turn a second-order kernel
into a fourth order kernel (\citealt{Schucany1977}, \citealt{Jones1993},
\citealt{Devroye1997}). For example, $K$ being a second-order kernel,
the kernels $\frac{4}{3}K(u)-\frac{1}{6}K\left(\frac{u}{2}\right)$,
$\frac{3}{2}K(u)+\frac{1}{2}uK^{'}(u)$, and $\frac{(s_{4}-s_{2}u^{2})K(u)}{s_{4}-s_{2}^{2}}$,
$s_{p}\triangleq\int_{\mathbb{R}}u^{p}K(u)du$ are known to be fourth-order
kernels, among many other examples. In the case of the (second-order)
Laplacian kernel $K(u)=\frac{1}{2}e^{-|u|}$ (equation \eqref{eq:Laplacian_kernel_1D}),
we obtain the following infinite-support fourth-order kernels:
\begin{align}
 & \frac{1}{3}\left(2e^{-\left|u\right|}-\frac{1}{4}e^{-\left|\frac{u}{2}\right|}\right)\label{eq:4th_order_lapA}\\
 & \frac{1}{4}\left(3-\left|u\right|\right)e^{-\left|u\right|}\label{eq:4th_order_lapB}\\
 & \frac{1}{5}\left(3-\frac{1}{4}u^{2}\right)e^{-\left|u\right|}\label{eq:4th_order_lapC}
\end{align}
which are all compatible with fast recursion (see the decompositions
of the similar kernels from subsection \ref{subsec:new-laplacian}
and Appendix \ref{sec:Matern32_kernel}). Beyond these simple examples,
the Laplacian kernel is also the root of the high-order class of Laguerre
kernels \citep{Berlinet1993}.

As pointed out previously, the fourth-order kernels \eqref{eq:4th_order_lapA}-\eqref{eq:4th_order_lapB}-\eqref{eq:4th_order_lapC}
necessarily take negative values, which can be deemed undesirable
in a variety of application contexts. Higher-order kernels can be
fixed to become non-negative (\citealt{Glad2003}, \citealt{Oudjane2005})
without loss of statistical performance, however the fast recursion
compatibility would be lost in the truncation process.

As a final remark, while there exists ``superkernels'' of infinite-order
(\citealt{Devroye1992}, \citealt{Politis1999}, \citealt{Hansen2005},
\citealt{Chacon2007}), to our knowledge none of them is compatible
with fast recursion.

\section{Numerical results\label{sec:numerics}}

Finally, this section reports numerical speed and accuracy results for multivariate CDF computation (subsection \ref{subsec:numeric_CDF}) and multivariate KDE computation (subsection \ref{subsec:numeric_KDE}).
Three approaches will be compared: 
\begin{itemize}
    \item the naive approach (direct computation of the sums \eqref{eq:ECDF} and \eqref{eq:Laplacian_kde_nD} independently for each evaluation point),
    \item the fast summation approach (subsection \ref{subsec:Fast-lexicographical-sweep}), and
    \item the fast divide-and-conquer approach  (subsection \ref{subsec:Fast-divide-and-conquer})
\end{itemize}
While the first approach is much slower than the other two, its results will serve as a benchmark for checking the accuracy of the other two methods.\\
Unless otherwise stated, we set the number of evaluation points $M$ to be equal to the number of input points $N$:
\begin{itemize}
    \item For the fast summation algorithm , we create an evaluation grid of shape $M_1\times M_2 \times \ldots \times M_d$ with $M_1=M_2=\ldots =M_d \triangleq N^{1/d}$, which ensures that $M=N$.
    \item For the fast divide-and-conquer algorithm , the evaluation points are equal to the input points, which also ensures that $M=N$.
    \item For the naive algorithm, we set the evaluation sample to the evaluation grid when comparing to the fast summation algorithm, and to the input points when comparing to the divide-and-conquer algorithm. 
\end{itemize}
The choices of input sample and bandwidth do not affect the speed or accuracy of the two proposed algorithms. For this reason, and for the sake of simplicity, we  arbitrarily choose to draw the $N$ input points from a $d$-dimensional Gaussian random variable
$X\sim\mathbb{N}(0,\mathbf{1_{d}})$  and to fix the bandwidth to $h=0.1$ in each dimension.\\
We perform the tests on an Intel\textregistered{} 
CPU i7-6820HQ @ 2.70GHz\footnote{https://ark.intel.com/content/www/fr/fr/ark/products/88970/intel-core-i7-6820hq-processor-8m-cache-up-to-3-60-ghz.html}.
The code was written in C++ and is available in the StOpt\footnote{https://gitlab.com/stochastic-control/StOpt}
library \citep{gevret2020stochastic}. Beyond CDF and KDE, StOpt implements fast kernel regression as well, as the weights $\omega_i$ in equation \eqref{eq:localdens} can be chosen in such a way as to cover all the terms needed to perform a Nadaraya-Watson kernel regression or a locally linear kernel regression (see for example Appendix B in \citealt{Langrene2019}).

\subsection{Cumulative distribution function\label{subsec:numeric_CDF}}

Table \ref{tab:2DCDFCalc} reports CDF calculation time (in seconds) on a bivariate example ($d=2$) with the naive, fast summation and divide-and-conquer approaches. We observe that, as expected, the fast summation and divide-and-conquer methods offer a massive speedup compared to naive summation (around 1 second for the fast algorithms vs. more than two hours for the direct computation for 1,28 million points for example), and both fast computation times are of the same order (as expected since $\mathcal{O}(N\log(N)^{(d-1){\vee}1})=\mathcal{O}(N{\log}N)$ when $d=2$).\\

\begin{table}[H]
\begin{centering}
\begin{tabular}{llllllll}
\toprule 
$\!\!\!$Nb particles & 20,000 & 40,000 & 80,000 & 160,000 & 320,000 & 640,000 & 1,280,000$\!\!\!$\tabularnewline
\midrule
$\!\!\!$Fast summation time & 0.01 & 0.01& 0.02& 0.04& 0.07& 0.15&  0.32  \tabularnewline
$\!\!\!$Divide-and-conquer time & 0.01& 0.02& 0.05 & 0.1 & 0.29 &0.66& 1.5 \tabularnewline
$\!\!\!$Naive time & 1.81 & 6.98 & 28 & 112 &  451 &  1939&  7586 \tabularnewline
\midrule 
\bottomrule
\end{tabular}
\par\end{centering}
\centering{}\caption{2D CDF calculation time (in seconds) \label{tab:2DCDFCalc}}
\end{table}
 
\begin{figure}[H]
\begin{minipage}[t]{0.48\columnwidth}%
\includegraphics[width=0.38\paperwidth]{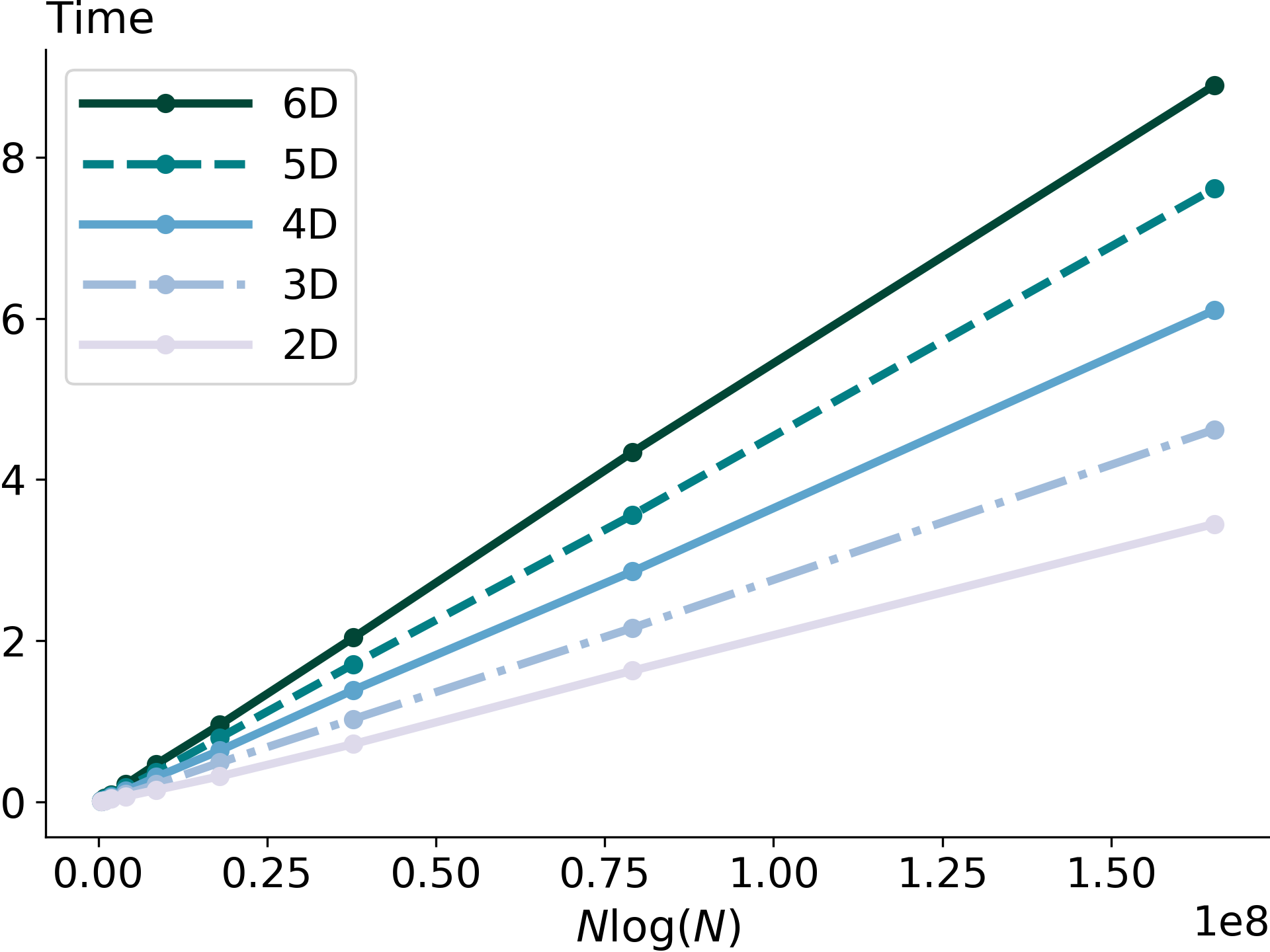}%
\end{minipage}\hfill{}%
\begin{minipage}[t]{0.48\columnwidth}%
\includegraphics[width=0.38\paperwidth]{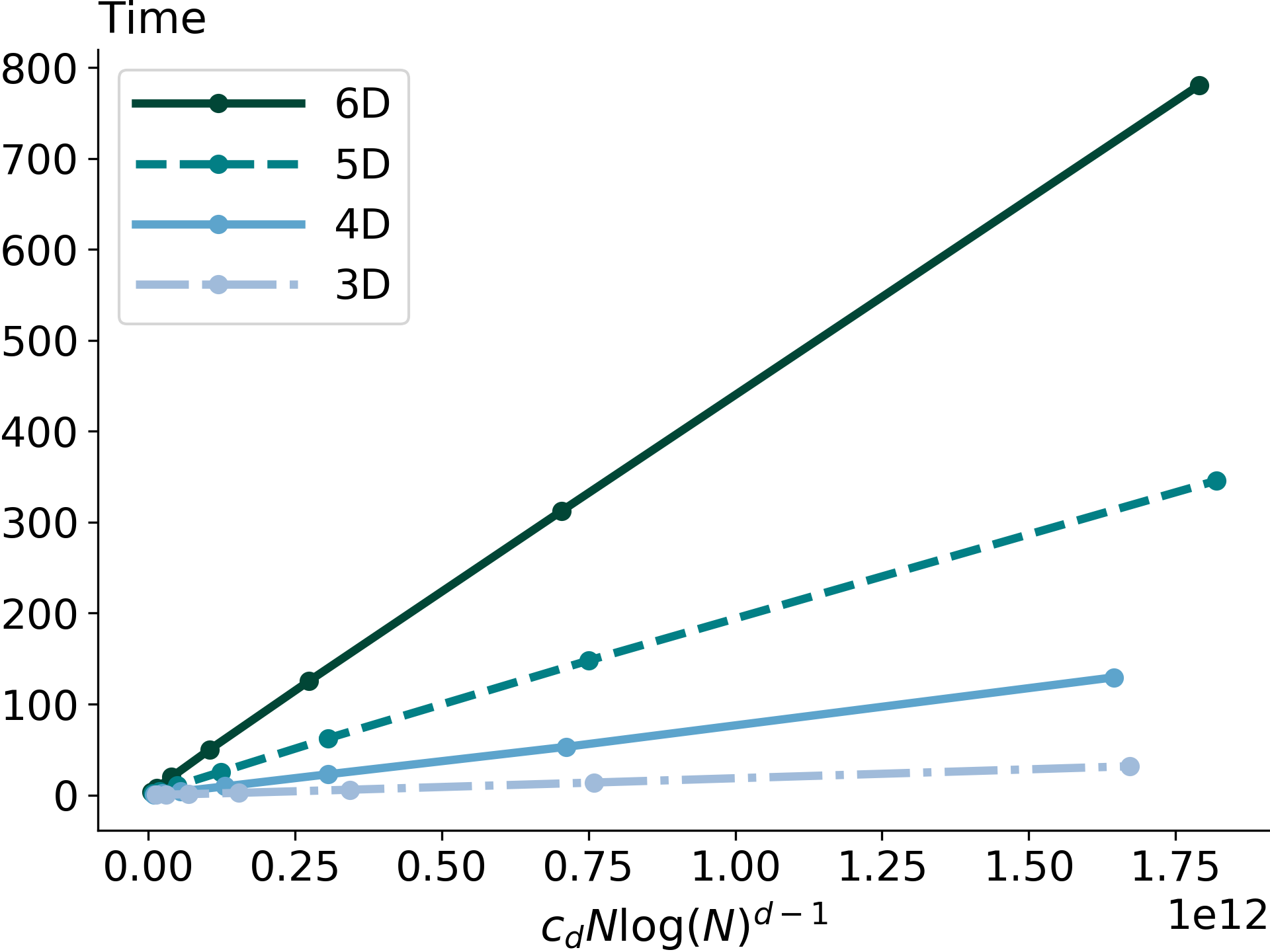}%
\end{minipage}
\caption{Runtime for CDF estimation (left: fast summation; right: divide-and-conquer)\label{fig:timeFastCDF}}
\end{figure}

 As the dimension increases, the computation time gap between divide-and-conquer and fast summation grows as expected, as shown on Table \ref{tab:6DCDFCalc}. 
 \begin{table}[H]
\begin{centering}
\begin{tabular}{llllllll}
\toprule 
$\!\!\!$Nb particles & 20,000 & 40,000 & 80,000 & 160,000 & 320,000 & 640,000 & 1,280,000$\!\!\!$\tabularnewline
\midrule
$\!\!\!$Fast summation time & 0.01& 0.02 & 0.05&  0.09& 0.22& 0.47& 0.96 \tabularnewline
$\!\!\!$Divide-and-conquer time & 1.2 &  3.1&  7.8&  19.7& 50.1&  125.1&  312.3  \tabularnewline
\midrule 
\bottomrule
\end{tabular}
\par\end{centering}
\centering{}\caption{6D CDF calculation time (in seconds) \label{tab:6DCDFCalc}}
\end{table}
 Figure \ref{fig:timeFastCDF} reports time calculation as a function of $N {\log}N$ for the fast summation approach and as a function of $c_d N \log(N)^{d-1}$ for the divide-and-conquer approach, with the scaling constants $c_3= 3000$, $c_4=200$, $c_5=15$, $c_6=1$ chosen to make the visual comparison easier. The resulting straight lines confirm the theoretical complexity.

\begin{figure}[H]
\begin{minipage}[t]{0.48\columnwidth}%
\includegraphics[width=\linewidth]{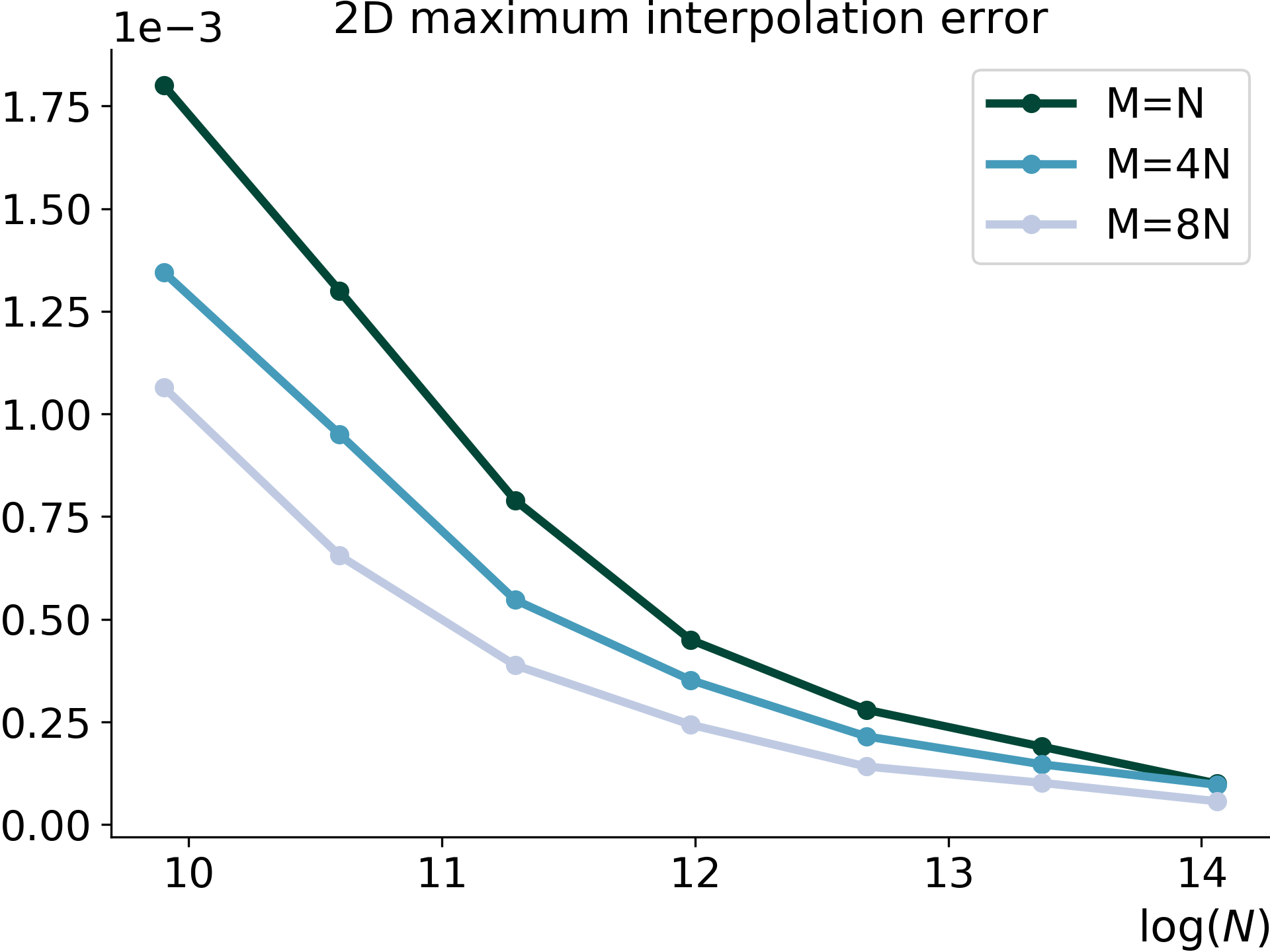}%
\end{minipage}
\begin{minipage}[t]{0.48\columnwidth}%
\includegraphics[width=\linewidth]{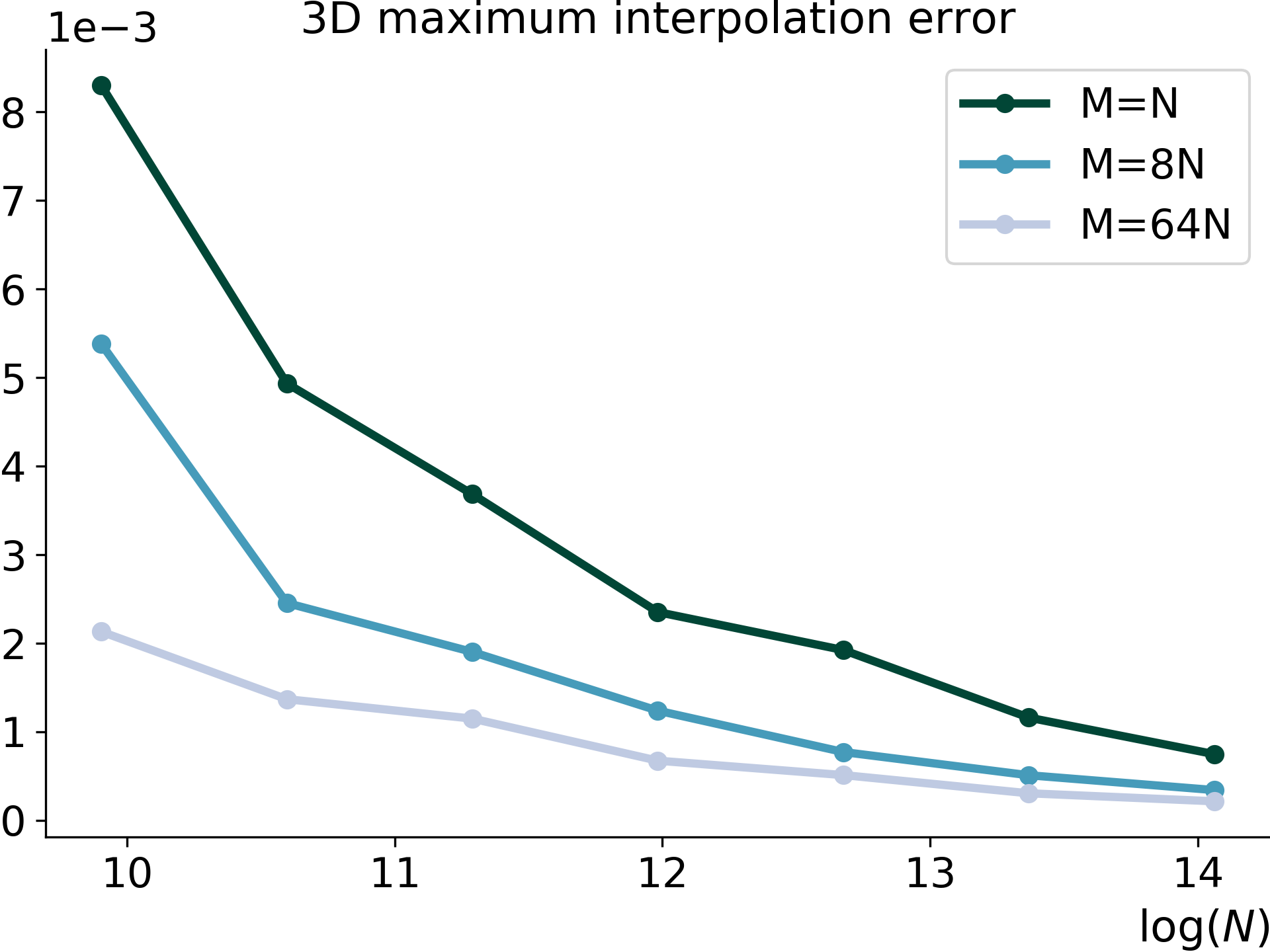}%
\end{minipage}
\begin{minipage}[t]{0.48\columnwidth}%
\includegraphics[width=\linewidth]{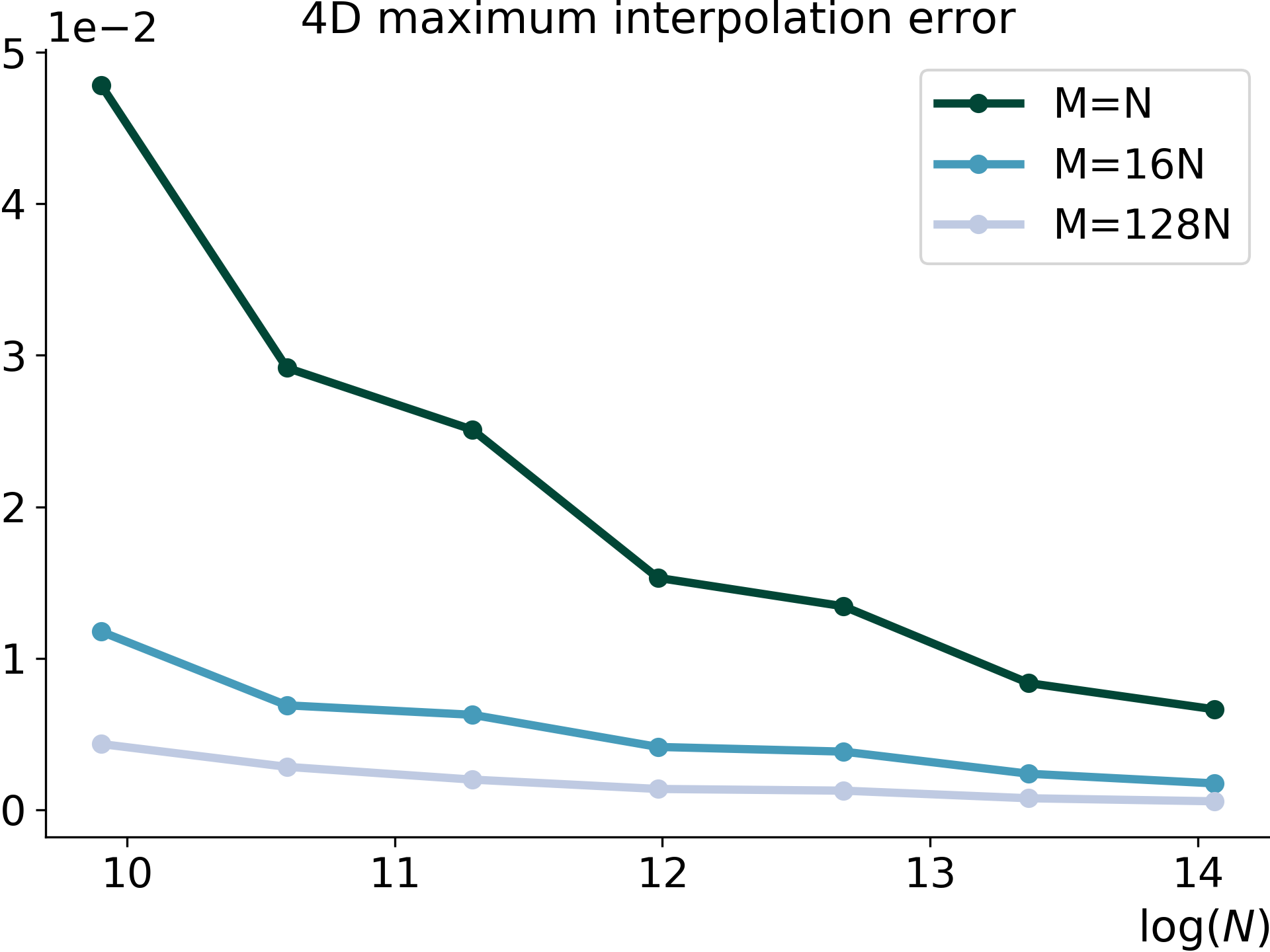}%
\end{minipage}
\begin{minipage}[t]{0.48\columnwidth}%
\includegraphics[width=\linewidth]{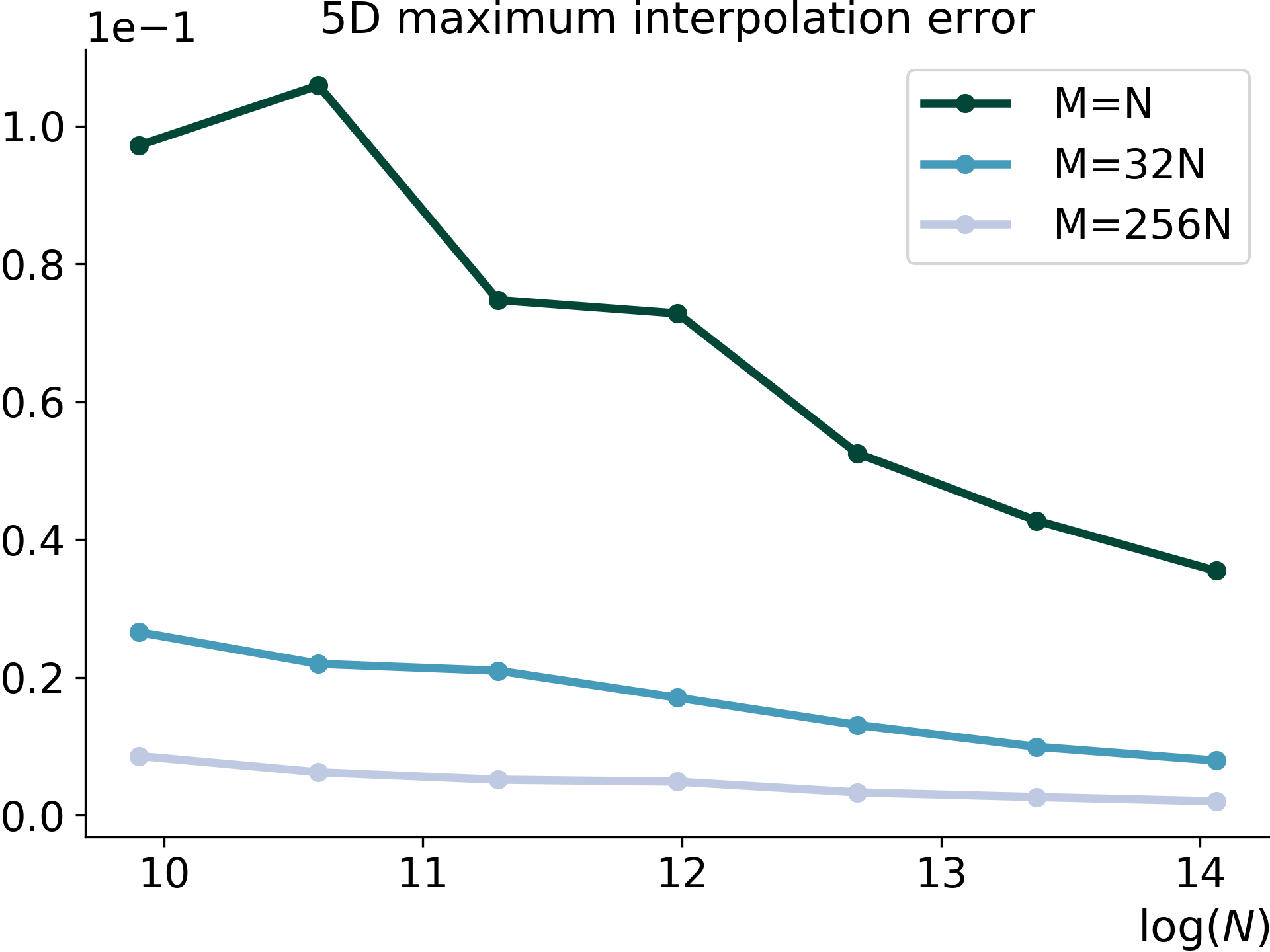}%
\end{minipage}
\caption{\label{fig:errorFastCDF}Maximal error on CDF for evaluation at sample points using  fast summation and interpolation taking as reference the divide and conquer calculation. }
\end{figure}

The CDF values calculated by the naive approach and the two fast methods are exactly the same with no rounding error whatsoever since the $y{\equiv}1$ case is a counting problem (integer count values with final division by $N$; see Remark \ref{rem:1N_scaling} on the fast summation case).\\
Suppose now that we specifically want to estimate the CDF values at the input points. The divide-and-conquer approach does this by design, while the fast summation approach requires an interpolation from the grid points to the input points. Figure \ref{fig:errorFastCDF} reports, for different numbers $M$ of evaluation points, the maximum interpolation error over the $N$ sample points between the CDF values computed by fast summation and linearly interpolated to the input points, and the divide-and-conquer CDF values (taken as reference).

When $M=N$, one can see that the worst-case interpolation error ranges from around 1\,${\scriptstyle \mathsf{E}}$-4 for $d=2$ and $N=1,280,000$ to around 1\,${\scriptstyle \mathsf{E}}$-1 for $d=5$ and $N=20,000$. This worst-case interpolation error is lower for small $d$ and large $N$, and can be reduced by using a finer evaluation grid, i.e. taking $M$ larger than $N$, as shown by the three curves on Figure \ref{fig:errorFastCDF}. Beyond linear interpolation, one could also resort to higher-order interpolation to reduce this error. Nevertheless, these results show that computing CDF values at input points by fast summation + interpolation is a viable method, with better results in the small $d$ high $N$ case.

\subsection{Kernel density estimation\label{subsec:numeric_KDE}}

We now perform the same numerical tests for kernel density estimation, more specifically Laplacian kernel density estimation (equation \eqref{eq:Laplacian_decomposition_nD}).\\
Table \ref{tab:2DKDECalc} reports KDE calculation time (in seconds) on a bivariate example with the naive, fast summation and divide-and-conquer approaches. Once again, the fast summation and divide-and-conquer methods offer a massive speedup compared to naive summation (respectively 0.34s and 2.29s vs. almost eight hours for the direct computation of \eqref{eq:Laplacian_kde_nD} for 0,64 million points for example), and both fast computation times are of the same order, up to a constant factor (around $6.0$).\\

\begin{figure}[H]
\begin{minipage}[t]{0.48\columnwidth}%
\includegraphics[width=0.38\paperwidth]{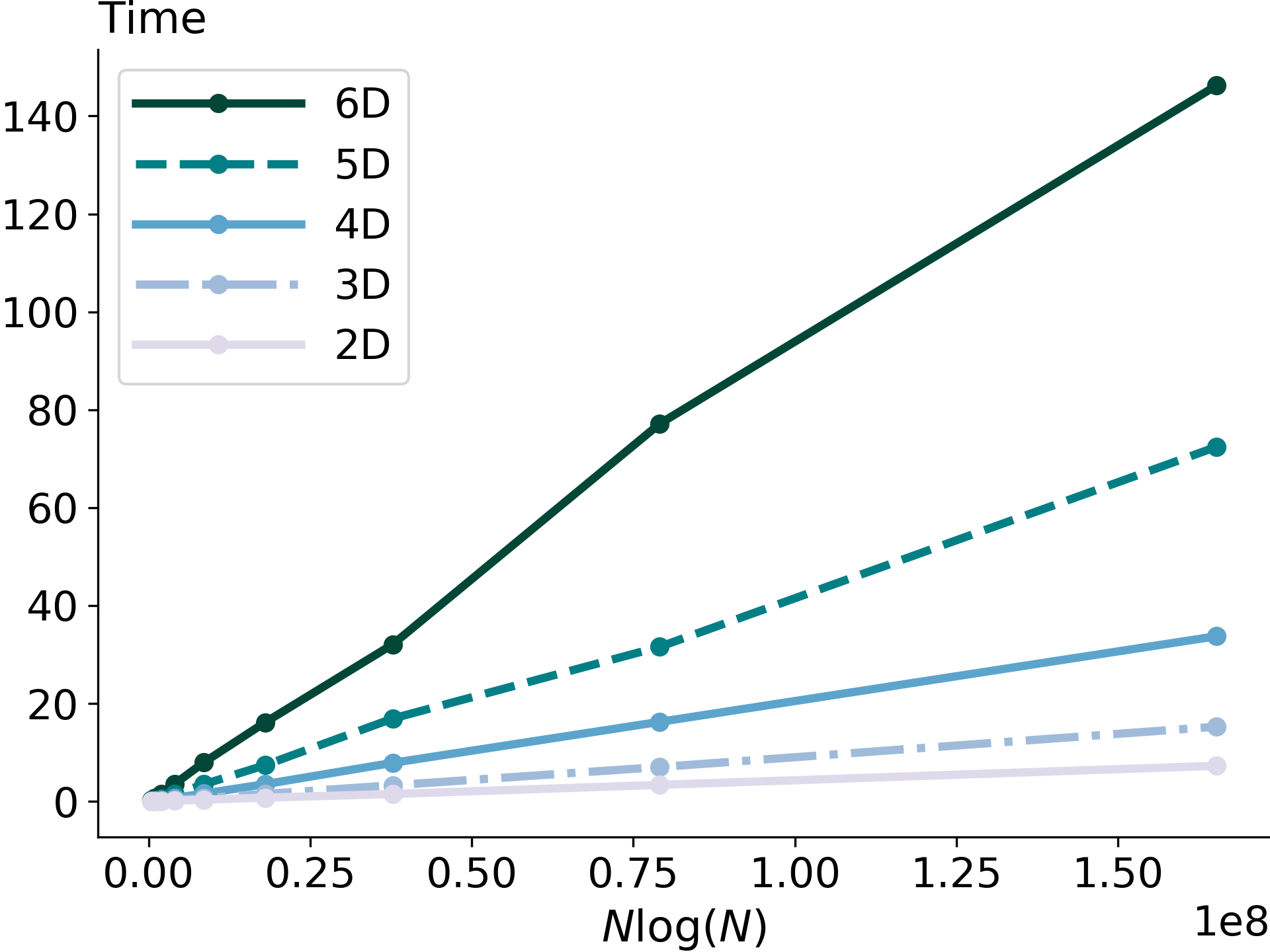}%
\end{minipage}\hfill{}%
\begin{minipage}[t]{0.48\columnwidth}%
\includegraphics[width=0.38\paperwidth]{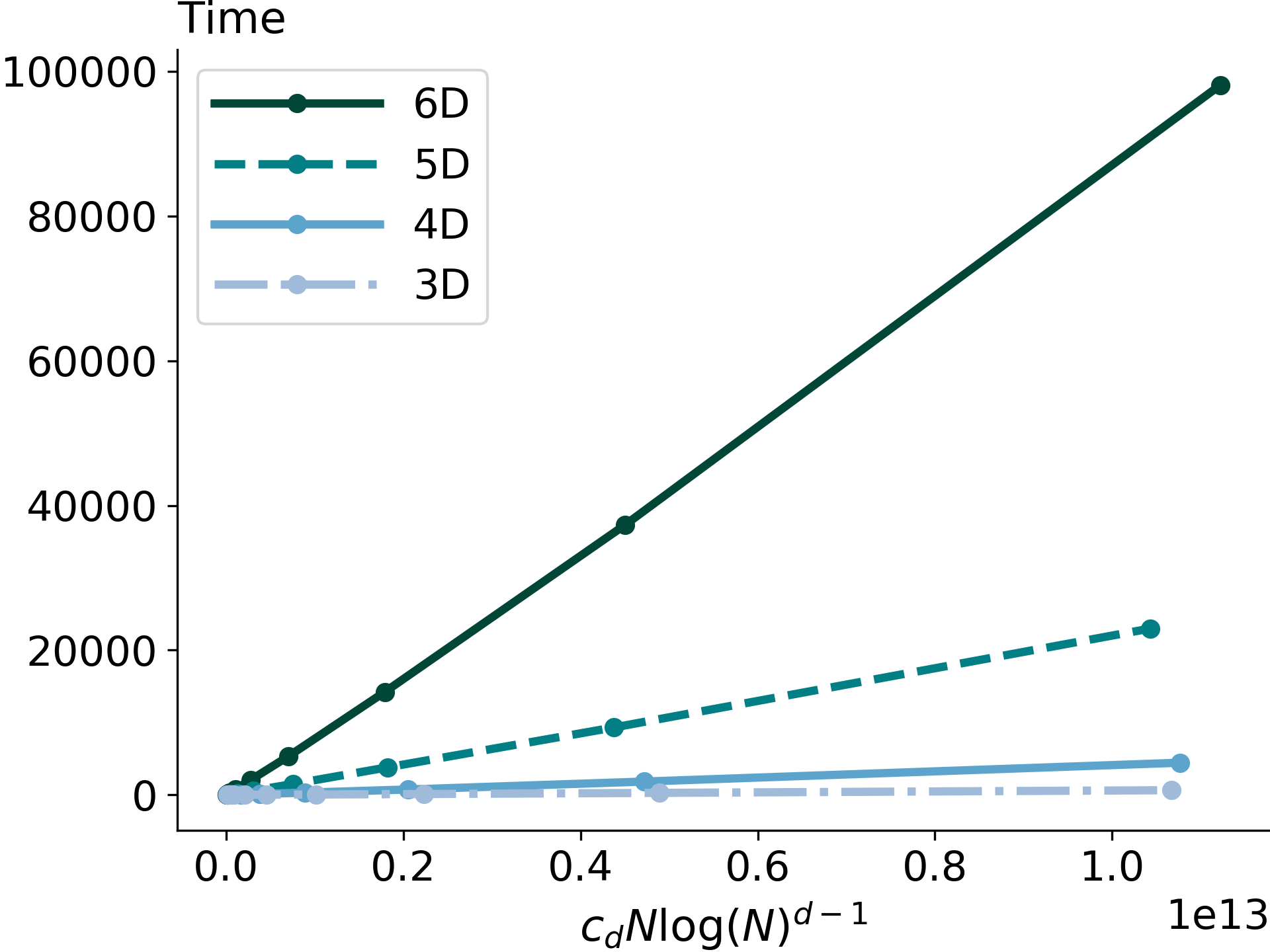}%
\end{minipage}
\caption{Runtime for KDE estimation (left: fast summation; right: divide-and-conquer)\label{fig:timeFastKDE}}
\end{figure}

\begin{table}[H]
\begin{centering}
\begin{tabular}{llllllll}
\toprule 
$\!\!\!$Nb particles & 20,000 & 40,000 & 80,000 & 160,000 & 320,000 & 640,000 &$\!\!\!$\tabularnewline
\midrule
$\!\!\!$Fast summation time &  0.01 & 0.01& 0.04& 0.08& 0.14 & 0.34 \tabularnewline
$\!\!\!$Divide-and-conquer time &  0.05 & 0.08 &   0.19& 0.43& 0.99& 2.29\tabularnewline
$\!\!\!$Naive time &  28 & 115 & 439 & 1742 & 7198 &  28132 \tabularnewline
\midrule 
\bottomrule
\end{tabular}
\par\end{centering}
\centering{}\caption{2D KDE calculation time (in seconds)\label{tab:2DKDECalc}}
\end{table}
As in the CDF case, the computation time gap between the two fast methods grows with the dimension, as shown on Table \ref{tab:6DKDECalc}.
\begin{table}[H]
\begin{centering}
\begin{tabular}{lllllllll}
\toprule 
$\!\!\!$Nb particles & 20,000 & 40,000 & 80,000 & 160,000 & 320,000 & 640,000 & 1,280,000 &$\!\!\!$\tabularnewline
\midrule
$\!\!\!$Fast summation time &  0.18&  0.26& 0.65& 1.51&  3.59& 7.97& 16.11 \tabularnewline
$\!\!\!$Divide-and-conquer time &  15 &41& 111& 294&  777 & 2040 & 5344\tabularnewline
\midrule 
\bottomrule
\end{tabular}
\par\end{centering}
\centering{}\caption{6D KDE calculation time (in seconds)\label{tab:6DKDECalc}}
\end{table}
 Figure \ref{fig:timeFastKDE} reports time calculation as a function of $N {\log}N$ for the fast summation approach and as a function of $c_d N \log(N)^{d-1}$ for the divide-and-conquer approach (with scaling constants $c_3= 4000$, $c_4=250$, $c_5=20$, $c_6=1$). Once again, the resulting straight lines confirm the theoretical complexity.

 As for accuracy, the maximum difference between the KDE values of the naive approach and those of both fast methods, caused by float rounding errors, remains below 1\,${\scriptstyle \mathsf{E}}$-14 independently of the dimension of the problem.\\
 
\begin{figure}[H]
\begin{minipage}[t]{0.48\columnwidth}%
\includegraphics[width=\linewidth]{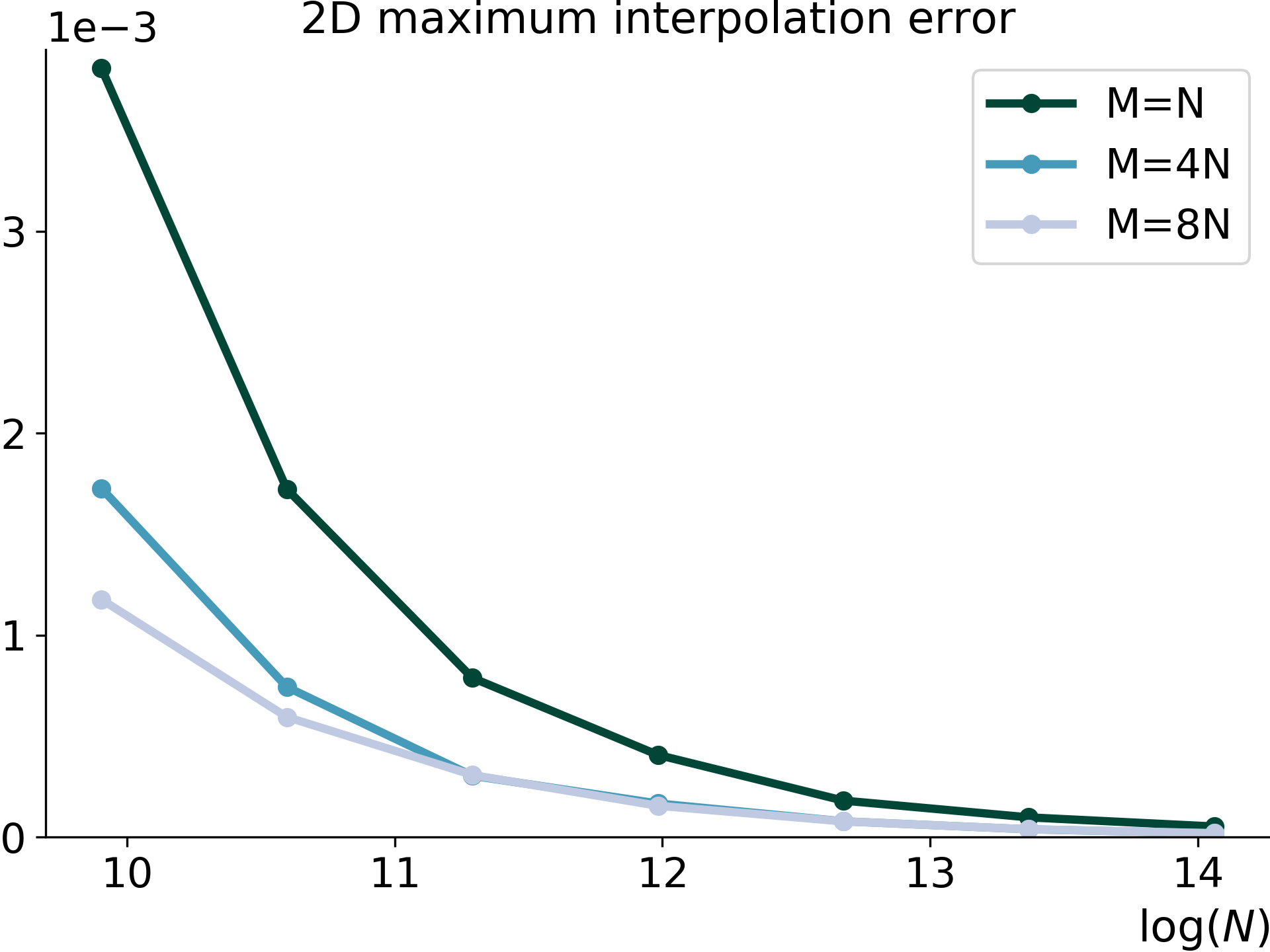}%
\end{minipage}
\begin{minipage}[t]{0.48\columnwidth}%
\includegraphics[width=\linewidth]{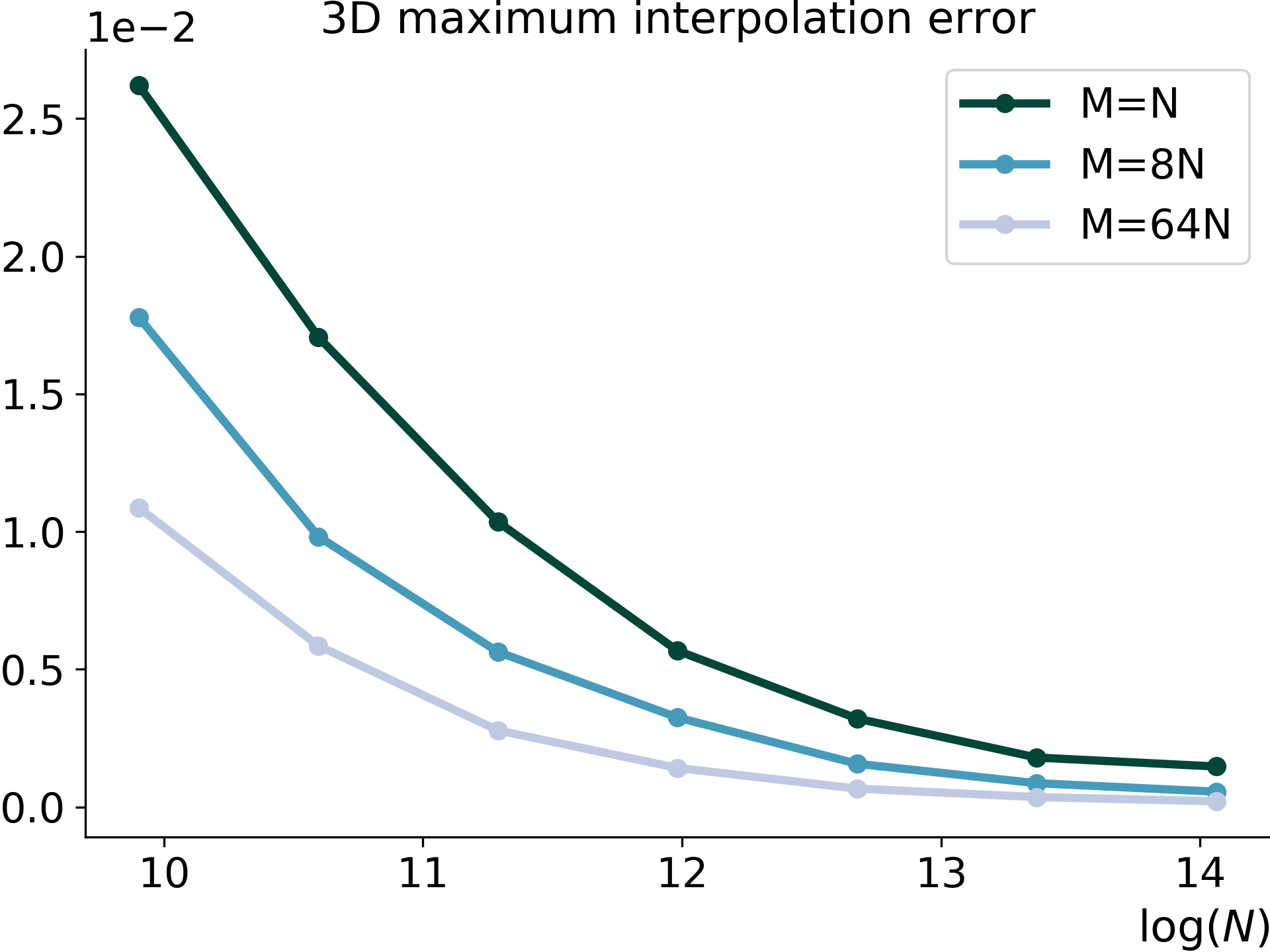}%
\end{minipage}
\begin{minipage}[t]{0.48\columnwidth}%
\includegraphics[width=\linewidth]{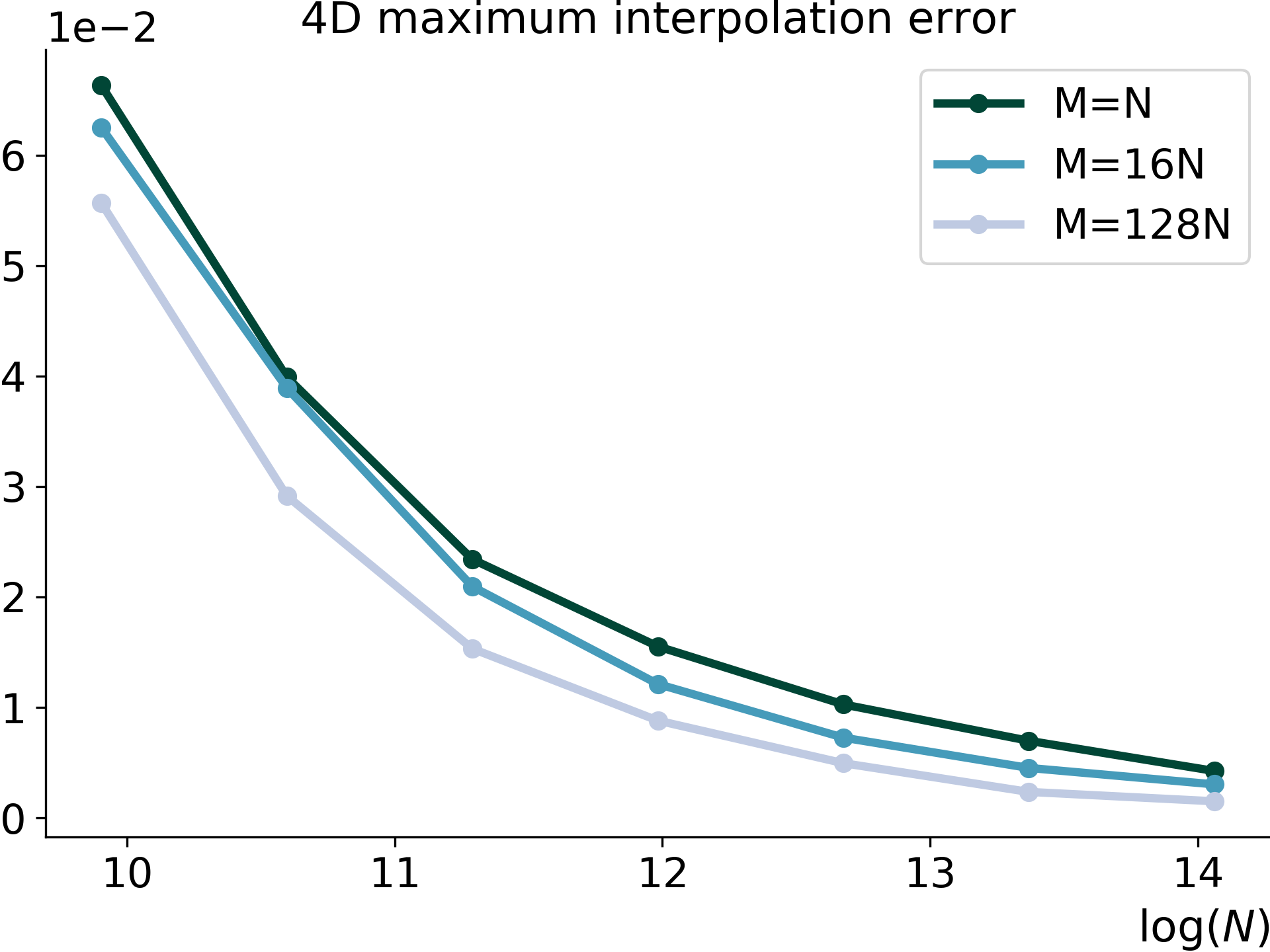}%
\end{minipage}
\begin{minipage}[t]{0.48\columnwidth}%
\includegraphics[width=\linewidth]{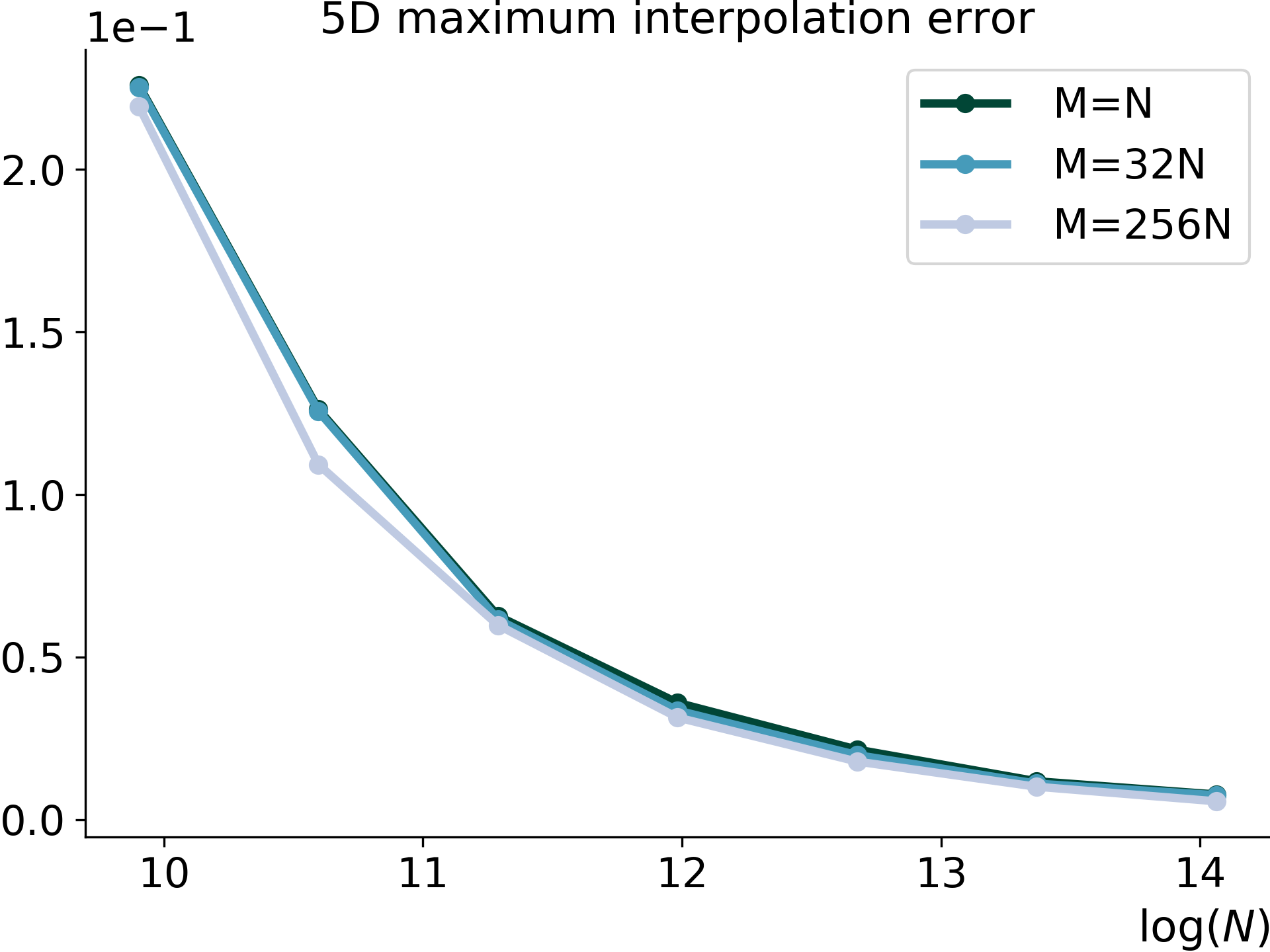}%
\end{minipage}
\caption{\label{fig:errorFastKDE}Maximal error on KDE evaluation at sample points using  fast summation and interpolation taking as reference the divide and conquer calculation. }
\end{figure}

 Finally, we also test the accuracy of the fast summation approach when the evaluation points are required to coincide with the input points, which requires an interpolation from the grid points. Figure \ref{fig:errorFastKDE} reports the maximum interpolation error over the $N$ sample points between the linearly interpolated CDF values computed by fast summation and the divide-and-conquer CDF values.
 
As in the CDF case, the worst-case interpolation error ranges between around 1\,${\scriptstyle \mathsf{E}}$-4 and 2\,${\scriptstyle \mathsf{E}}$-1, is smaller for small $d$, large $N$ or large $M$. However,  the accuracy improvements obtained by increasing $M$ get smaller in higher dimension. Nevertheless, the fast summation + interpolation approach can still be considered a viable option for KDE estimation at the input data points, provided $d$ is small or $N$ is large.

\section{Conclusion\label{sec:conclusion}}
A new algorithm based on fast summation in lexicographical order has been developed to efficiently calculate multivariate empirical cumulative distribution functions (ECDFs)  with $O(N{\log}N)$ computational cost for $N$ arbitrary data points and $N$ evaluation points on a rectilinear grid. Numerical tests and comparisons to a state-of-the-art $\mathcal{O}(N\log(N)^{(d-1){\vee}1})$ divide-and-conquer algorithm confirm the speed of this exact algorithm.\\
Besides, we establish a multivariate decomposition formula of kernel density estimators (KDEs) into a weighted sum of generalized ECDFs for a large class of kernels. This connection leads to new fast KDE algorithms: one based on fast summation with $O(N{\log}N)$ complexity, and one based on divide-and-conquer recursion with $\mathcal{O}(N\log(N)^{(d-1){\vee}1})$ complexity.\\
The class of compatible kernels includes classical kernels such as the uniform, Epanechnikov and Laplacian kernels. We show that it also includes the Sargan and Mat\'ern kernels, which can be used to approximate incompatible kernels such as the Gaussian kernel.\\
Following our computational breakthrough, several possible extensions and potential future work come to mind:
\begin{itemize}
\item The investigation of computational methods for the related kernel \emph{distribution} estimation problem (\citealt{Yamato1973}, \citealt{Liu2008}) based on the algorithmic approaches developed in this paper.
\item Further investigation of the promising class of multivariate polynomial-exponential kernels, in particular their ability to approximate multivariate kernels, and their ability to speed up statistical techniques based on multivariate Gaussian variables using the fast algorithms from this paper.
\item The application of fast kernel regression for image processing, as uniform pixel grids are an ideal ground for Algorithm \ref{algo:fast_sum_CDF} for which its computational complexity is an optimal $\mathcal{O}(N)$.
\item The comparison, more generally, of our algorithms to fast convolution methods such as the Fast Fourier Transform (FFT) for compatible convolution kernels.
\end{itemize}

\bibliographystyle{apacite}
\bibliography{jobname}

\clearpage

\appendix

\section{Computation of local sums\label{sec:local-sums}}

This appendix details how to efficiently compute the local sums $s_{j_{1},j_{2},\ldots,j_{d}}$
defined in equation \eqref{eq:partial_sums} (subsection \ref{subsec:Fast-lexicographical-sweep}).
Subsection \ref{subsec:General-case} details the general case, based
on independent input data sorting in each dimension, for a $\mathcal{O}(N\log N)$
computational cost. Subsection \ref{subsec:Uniform-grid-case} details
the uniform grid case: in this special case, the computational cost
can be brought down to $\mathcal{O}(N)$ by using constant mesh divisions
as a substitute to sorting.

\subsection{General case\label{subsec:General-case}}

\begin{algorithm2e}[H]
\DontPrintSemicolon 
\SetAlgoLined 

\vspace{1.5mm}

\KwIn{input sample $x_{i}=(x_{1,i},\ldots,x_{d,i})$, $i=1,2,\ldots,N$}

\vspace{0.5mm}

\KwIn{evaluation grid $(z_{1,j_{1}},z_{2,j_{2}},\ldots,z_{d,j_{d}})$,
$j_{k}\in\{1,2,\ldots,M_{k}\}$, $k\in\{1,2,\ldots,d\}$}

\vspace{1mm}

Define index matrix $\mathrm{INDEX}[k,i]$ \Comment*[l]{local sum
index $\in\{1,2,M_{k}+1\}$}

where $k=1,2,\ldots,d$ and $i=1,2,\ldots,N$

\For{$\textcolor{blue}{\mathbf{(}}$ $k=1,2,...\hspace{0.0833em},d$
$\textcolor{blue}{\mathbf{)}}$}{\vspace{0.5mm}

Sort the set $\left\{ x_{k,1},\ldots x_{k,N}\right\} $ in increasing
order, using for example quicksort or

mergesort ($\mathcal{O}(N\log N)$): define the permutation $\phi_{k}:\{1,2,\ldots,N\}\mapsto\{1,2,\ldots,N\}$

such that\vspace{-0.5mm}
\begin{equation}
x_{k,\phi_{k}(1)}<x_{k,\phi_{k}(2)}<\cdots<x_{k,\phi_{k}(N)}\label{eq:sorted_kdim}
\end{equation}

$x_{\mathrm{idx}}=1$ \Comment*[l]{input index $\in\{1,2,\ldots,N\}$}

$z_{\mathrm{idx}}=1$ \Comment*[l]{evaluation grid index $\in\{1,2,\ldots,M_{k}\}$}\vspace{1mm}

\While{$\textcolor{blue}{\mathbf{(}}$ $x_{\mathrm{idx}}\leq N$
$\textcolor{blue}{\mathbf{)}}$}{

\eIf{$\textcolor{blue}{\mathbf{(}}$ $x_{k,\phi_{k}(x_{\mathrm{idx}})}\leq z_{k,z_{\mathrm{idx}}}$
$\textcolor{blue}{\mathbf{)}}$}{\vspace{2.5mm}

$\mathrm{INDEX}[k,\phi_{k}(x_{\mathrm{idx}})]=z_{\mathrm{idx}}$

$x_{\mathrm{idx}}\,+\hspace{-0.0833em}\!=1$

}{

$z_{\mathrm{idx}}\,+\hspace{-0.0833em}\!=1$

}

}

}

$s_{j_{1},j_{2},\ldots,j_{d}}=0$, ${\scriptstyle \forall(j_{1},j_{2},\ldots,j_{d})\in\{1,2,\ldots,M_{1}+1\}\times\ldots\times\{1,2,\ldots,M_{d}+1\}}$\vspace{1mm}

\For{$\textcolor{blue}{\mathbf{(}}$ $i=1,2,...\hspace{0.0833em},N$
$\textcolor{blue}{\mathbf{)}}$}{

$s_{\mathrm{INDEX}[1,i],\,\mathrm{INDEX}[2,i],\,\ldots,\,\mathrm{INDEX}[d,i]}\,+\hspace{-0.0833em}\!=y_{i}/N$

}

\KwOut{$s_{j_{1},j_{2},\ldots,j_{d}}=\frac{1}{N}\sum_{i=1}^{N}y_{i}\mathbbm{1}\{z_{1,j_{1}-1}<x_{1,i}\leq z_{1,j_{1}},\ldots,z_{d,j_{d}-1}<x_{d,i}\leq z_{d,j_{d}}\}$

\vspace{1.5mm}

\hspace{17mm}for every local sum index ${\scriptstyle (j_{1},j_{2},\ldots,j_{d})\in\{1,2,\ldots,M_{1}+1\}\times\ldots\times\{1,2,\ldots,M_{d}+1\}}$

}

\vspace{1mm}

\caption{Fast computation of local sums by independent sorting in
each dimension\label{algo:local_sum_general}}

\end{algorithm2e}

Algorithm \ref{algo:local_sum_general} has a $\mathcal{O}(N{\log}N)$
computational complexity, owing to the data sorting in each dimension.
Its memory complexity is $\mathcal{O}(N+M)$.

\begin{rem}
An alternative algorithm to compute the same local sums has been proposed in \citet{bouchard2012monte}. It is based on partial sorts in each dimension and its computational complexity is $\mathcal{O}((\sum_{k=1}^d M_k+1)N)$. This complexity is better than $\mathcal{O}(N\log N)$ when $M{\ll}{\log}(N)^d$. However, in the case when $M{\approx}N$ (and $M_1=M_2=\ldots=M_d$), its equivalent $\mathcal{O}(N^{1+\frac{1}{d}})$ complexity does not improve over Algorithm \ref{algo:local_sum_general}. 
\end{rem}

\subsection{Uniform grid case\label{subsec:Uniform-grid-case}}

\begin{algorithm2e}[H]
\DontPrintSemicolon 
\SetAlgoLined 

\vspace{1.5mm}

\KwIn{input sample $x_{i}=(x_{1,i},\ldots,x_{d,i})$, $i=1,2,\ldots,N$}

\vspace{0.5mm}

\KwIn{evaluation grid $(z_{1,j_{1}},z_{2,j_{2}},\ldots,z_{d,j_{d}})$,
$j_{k}\in\{1,2,\ldots,M_{k}\}$, $k\in\{1,2,\ldots,d\}$}

\vspace{1mm}

Define index matrix $\mathrm{INDEX}[k,i]$ \Comment*[l]{local sum
index $\in\{1,2,M_{k}+1\}$}

where $k=1,2,\ldots,d$ and $i=1,2,\ldots,N$

\For{$\textcolor{blue}{\mathbf{(}}$ $k=1,2,...\hspace{0.0833em},d$
$\textcolor{blue}{\mathbf{)}}$}{\vspace{0.5mm}

$\Delta z_{k}\triangleq z_{k,2}-z_{k,1}$ \Comment*[l]{constant mesh
$=z_{k,3}-z_{k,2}=z_{k,4}-z_{k,3}=\ldots$}

\vspace{1mm}

\For{$\textcolor{blue}{\mathbf{(}}$ $i=1,2,...\hspace{0.0833em},N$
$\textcolor{blue}{\mathbf{)}}$}{

\Comment*[l]{mesh division rounded to upper integer}

$\mathrm{INDEX}[k,i]=\max(\,1\,,\,\min(\,M_{k}+1\,,\,1+\left\lceil(x_{k,i}-z_{k,1})/\Delta z_{k}\right\rceil\,)\,)$ 

}

}

$s_{j_{1},j_{2},\ldots,j_{d}}=0$, ${\scriptstyle \forall(j_{1},j_{2},\ldots,j_{d})\in\{1,2,\ldots,M_{1}+1\}\times\ldots\times\{1,2,\ldots,M_{d}+1\}}$\vspace{1mm}

\For{$\textcolor{blue}{\mathbf{(}}$ $i=1,2,...\hspace{0.0833em},N$
$\textcolor{blue}{\mathbf{)}}$}{

$s_{\mathrm{INDEX}[1,i],\,\mathrm{INDEX}[2,i],\,\ldots,\,\mathrm{INDEX}[d,i]}\,+\hspace{-0.0833em}\!=y_{i}/N$

}

\KwOut{$s_{j_{1},j_{2},\ldots,j_{d}}=\frac{1}{N}\sum_{i=1}^{N}y_{i}\mathbbm{1}\{z_{1,j_{1}-1}<x_{1,i}\leq z_{1,j_{1}},\ldots,z_{d,j_{d}-1}<x_{d,i}\leq z_{d,j_{d}}\}$

\vspace{1.5mm}

\hspace{17mm}for every local sum index ${\scriptstyle (j_{1},j_{2},\ldots,j_{d})\in\{1,2,\ldots,M_{1}+1\}\times\ldots\times\{1,2,\ldots,M_{d}+1\}}$

}

\vspace{1mm}

\caption{Fast computation of local sums by mesh division on uniform
grid\label{algo:local_sum_uniform}}

\end{algorithm2e}

Algorithm \ref{algo:local_sum_uniform} has a $\mathcal{O}(N)$ computational
complexity, and $\mathcal{O}(N+M)$ memory complexity.

\section{General matrix bandwidth\label{sec:General-matrix-bandwidth}}

The general multivariate weighted Parzen-Rosenblatt kernel density
estimator is defined by: 

\begin{equation}
\hat{f}_{\mathrm{KDE}}(z)=\frac{1}{\left|H\right|^{1/2}N}\sum_{i=1}^{N}w_{i}K_{\hspace{-0.0833em}d}\!\left(H^{-1/2}(x_{i}-z)\right)\label{eq:kde_nD_H}
\end{equation}
where $H$ is a symmetric positive definite $d\times d$ bandwidth
matrix, see for example \citet{Wand1995}. As pointed out in \citet{Langrene2019},
one can without loss of generality focus on the diagonal bandwidth
case $H=\mathrm{diag}(h)$, where $h=(h_{1},h_{2},\ldots,h_{d})\in\mathbb{R}^{d}$.
Indeed, the eigenvalue decomposition of the symmetric positive definite
matrix $H$ is given by $H=R\Delta^{2}R^{\top}$ where $R$ is a rotation
matrix and $\Delta=\mathrm{diag}(h)$ is a diagonal matrix with strictly
positive diagonal elements. Consequently, $H^{-1/2}(x_{i}-z)=\mathrm{diag}(\frac{1}{h})(R^{T}x_{i}-R^{T}z)$.
By rotating both the input points $x_{i}$ and the evaluation point
$z$ using the rotation matrix $R^{T}$, the multivariate kernel density
estimator \eqref{eq:kde_nD_H} becomes
\begin{equation}
\hat{f}_{\mathrm{KDE}}(z)=\frac{1}{N\Pi_{k=1}^{d}h_{k}}\sum_{i=1}^{N}w_{i}K_{\hspace{-0.0833em}d}\!\left(\frac{x_{i}-z}{h}\right)\label{eq:kde_nD}
\end{equation}
where $x_{i}$ and $z$ denote respectively the input points and evaluation
point in the new coordinates. In the Laplacian kernel case, equation
\eqref{eq:kde_nD} turns into the multivariate KDE equation \eqref{eq:Laplacian_kde_nD}
used in Section \ref{sec:Fast-KDE}.

\section{Multivariate Mat\'ern kernel\label{sec:Matern32_kernel}}

Taking the Mat\'ern-3/2 kernel \eqref{eq:Matern32} ($p=1$) with
$h=\sqrt{3}$ as example, several approaches exist to define a multivariate
kernel. One approach, known as product kernel, is to multiply univariate
kernels:
\begin{equation}
K_{\hspace{-0.0833em}d}(u)=\frac{1}{4^{d}}\prod_{k=1}^{d}\left(1+\left|u_{k}\right|\right)e^{-\left|u_{k}\right|}\label{eq:Matern32_product}
\end{equation}
Another approach is to replace the absolute value $\left|u\right|$
by the L1 norm $\left\Vert u\right\Vert _{1}=\sum_{k=1}^{d}\left|u_{k}\right|$,
along with a correction of the normalization constant:
\begin{equation}
K_{\hspace{-0.0833em}d}(u)=\frac{1}{2^{d}(1+d)}\left(1+\sum_{k=1}^{d}\left|u_{k}\right|\right)e^{-\sum_{k=1}^{d}\left|u_{k}\right|}\label{eq:Matern32_average}
\end{equation}

The product approach \eqref{eq:Matern32_product} preserves the continuous
differentiability of the kernel, which is not the case for the additive
approach \eqref{eq:Matern32_average}. Nevertheless, the CDF decomposition
of the additive kernel \eqref{eq:Matern32_average} contains significantly
fewer terms than the one of the product kernel \eqref{eq:Matern32_product}.
Indeed, the KDE decomposition of \eqref{eq:Matern32_average} is given
by

\begin{align}
 & \frac{1}{N\prod_{k=1}^{d}h_{k}}\sum_{i=1}^{N}w_{i}K_{\hspace{-0.0833em}d}\left(\frac{x_{i}-z}{h}\right)\nonumber \\
= & \frac{1}{2^{d}(1+d)}\frac{1}{N\prod_{k=1}^{d}h_{k}}\sum_{i=1}^{N}w_{i}\left(1+\sum_{l=1}^{d}\left|\frac{x_{l,i}-z_{l}}{h_{l}}\right|\right)e^{-\sum_{k=1}^{d}\left|\frac{x_{k,i}-z_{k}}{h_{k}}\right|}\nonumber \\
= & \frac{1}{2^{d}(1+d)}\frac{1}{N\prod_{k=1}^{d}h_{k}}\sum_{i=1}^{N}w_{i}\left(1+\sum_{l=1}^{d}\left|\frac{x_{l,i}-z_{l}}{h_{l}}\right|\right)\nonumber \\
 & \left.\times\prod_{k=1}^{d}\left(e^{-\frac{z_{k}}{h_{k}}}e^{\frac{x_{k,i}}{h_{k}}}\mathbbm{1}\{x_{k,i}\leq z_{k}\}+e^{\frac{z_{k}}{h_{k}}}e^{-\frac{x_{k,i}}{h_{k}}}\mathbbm{1}\{-x_{k,i}<-z_{k}\}\right)\right)\nonumber \\
= & \frac{1}{2^{d}(1+d)}\frac{1}{N\prod_{k=1}^{d}h_{k}}\sum_{i=1}^{N}\sum_{\delta\in\{-1,1\}^{d}}e^{-\sum_{k=1}^{d}\frac{\delta_{k}z_{k}}{h_{k}}}\left(1+\sum_{l=1}^{d}\left|\frac{x_{l,i}-z_{l}}{h_{l}}\right|\right)\nonumber \\
 & \left.\times w_{i}e^{\sum_{k=1}^{d}\frac{\delta_{k}x_{k,i}}{h_{k}}}\mathbbm{1}\{\delta_{1}x_{1,i}\leq_{\delta_{1}}\delta_{1}z_{1},\ldots,\delta_{d}x_{d,i}\leq_{\delta_{d}}\delta_{d}z_{d}\}\right)\nonumber \\
= & \frac{1}{2^{d}(1+d)}\frac{1}{N\prod_{k=1}^{d}h_{k}}\sum_{i=1}^{N}\sum_{\delta\in\{-1,1\}^{d}}e^{-\sum_{k=1}^{d}\frac{\delta_{k}z_{k}}{h_{k}}}\left(1-\sum_{l=1}^{d}\delta_{l}\frac{x_{l,i}-z_{l}}{h_{l}}\right)\nonumber \\
 & \left.\times w_{i}e^{\sum_{k=1}^{d}\frac{\delta_{k}x_{k,i}}{h_{k}}}\mathbbm{1}\{\delta_{1}x_{1,i}\leq_{\delta_{1}}\delta_{1}z_{1},\ldots,\delta_{d}x_{d,i}\leq_{\delta_{d}}\delta_{d}z_{d}\}\right)\nonumber \\
= & \frac{1}{2^{d}(1+d)}\frac{1}{\prod_{k=1}^{d}h_{k}}\sum_{\delta\in\{-1,1\}^{d}}e^{-\sum_{k=1}^{d}\frac{\delta_{k}z_{k}}{h_{k}}}\left(\left(1+\sum_{l=1}^{d}\frac{\delta_{l}z_{l}}{h_{l}}\right)F_{N}(\delta z,\delta;\delta x,y^{(0)})-\sum_{l=1}^{d}\frac{\delta_{l}}{h_{l}}F_{N}(\delta z,\delta;\delta x,y^{(l)})\right)\label{eq:Matern32_decomposition_nD}
\end{align}
with $y_{i}^{(0)}=y_{i}^{(0)}(\delta)\triangleq w_{i}e^{\sum_{k=1}^{d}\frac{\delta_{k}x_{k,i}}{h_{k}}}$
and $y_{i}^{(l)}=y_{i}^{(l)}(\delta)\triangleq w_{i}x_{l,i}e^{\sum_{k=1}^{d}\frac{\delta_{k}x_{k,i}}{h_{k}}}$.
This decomposition contains $2^{d}(d+1)$ CDFs to compute. By contrast,
similar computations show that the KDE decomposition of the product
kernel \eqref{eq:Matern32_product} contains a total of $4^{d}$ CDFs
to compute. In other words, the additive kernel \eqref{eq:Matern32_average}
is much more attractive than the product kernel \eqref{eq:Matern32_product}
from a computational point of view, even when accounting for its lower
efficiency. These two kernels are however not as computationally attractive
as the Laplacian and uniform kernels, whose CDF decompositions contain
$2^{d}$ terms (\eqref{eq:Laplacian_decomposition_nD} and \eqref{eq:uniform_decomposition_nD}).

\section{Divide-and-conquer for Laplacian kernel density estimation}

This Appendix explains how to adapt the divide-and-conquer algorithm described in Section \ref{subsec:Fast-divide-and-conquer}  to compute the $2^{d}$ CDF vectors $\{F_{N}(x_{i},\delta)\}_{i=1,N}$ required to compute equation \eqref{eq:Laplacian_decomposition_nD}.
A possible approach would consist in adapting the algorithm \ref{sec:FastCDF} used to calculate \eqref{eq:ECDFdelta} with $\delta = (1,..,1)$ by applying a modified version
$2^{d}$ times to calculate the different terms.\\
We propose a single algorithm, implemented in the StOpt library, which makes it possible to compute the $F_N$ for all the $\delta$ in one recursion, avoiding to sort the particles $2^d$ times.\\
We give the algorithm obtained in general dimension to calculate for all  $\delta \in \{-1,1\}^{d}$, and given $m$, $l$ with values in $1, \dots, d$, $(p,q) \in \mathds{N}^2$  a general term for $j=1,N$  $$\sum_{i=1}^N x_{l,i}^{p} x_{m,i}^{q} e^{\sum_{k=1}^{d}\frac{\delta_{k}x_{k,i}}{h_{k}}} \mathbbm{1}\{\delta_{1}x_{1,i}<\delta_{1}x_{1,j},\ldots,\delta_{d}x_{d,i}<\delta_{d}x_{d,j}\}.$$
Once again observe that the inequalities are strict in the expression above. As before the tests for non-empty sets are dropped out.
\begin{itemize}
    \item Algorithm \ref{algo:divideMain} is the main calling  similar to \ref{algo:divideMainECDF}.
    The special 2D case is dealt with the call of the two different one-dimensional merge algorithm \ref{algo:Merge1D1} and \ref{algo:Merge1D2} instead of a single one-dimensional algorithm.
    \item The $n$-dimensional merge algorithm \ref{algo:MergeND} is similar to Algorithm \ref{algo:MergeNDECDF}.  Besides, a set $\Delta$ of $\delta \in \{-1,1\}^d$ is given as input too such that either $\delta_k x_k \le \delta_k y_k$ for $ k > I_{dim}$ for all $x \in \kappa_1$ and $y \in \kappa_2$ or   $\delta_k x_k >  \delta_k y_k$ for $ k > I_{dim}$ for all $x \in \kappa_1$ and $y \in \kappa_2$.\\
    For the couple of sets $(\kappa_{1,1},\kappa_{2,2})$,$(\kappa_{2,1},\kappa_{1,2})$ where dominance is clear in the current dimension, the $n$-dimensional merge algorithm is called in the dimension below and some subset of $\Delta$. In the case when $I_{dim}=2$, a direct call to the one-dimensional merge algorithms \ref{algo:Merge1D1} and \ref{algo:Merge1D2} is performed. 
    \item Two one-dimensional merge in dimension 1  are used. The first version $\mathbf{Merge1D1}$ is used for the $\delta$ such that $\delta_1=1$ and is the same as the $\mathbf{Merge1D}$ algorithm except that it works for a set of $\delta$ given as input. The second one is for the  $\delta$ such that $\delta_1=-1$. 
\end{itemize}

\begin{algorithm2e}[H]
\DontPrintSemicolon 
\SetAlgoLined 
\vspace{1mm}
\KwIn{ $x = (x_i,  \dots, x_N)$, $\psi(x_i,\delta) =  x_{l,i}^{p} x_{m,i}^{q} e^{-\sum_{k=1}^{d}\frac{\delta_{k}x_{k,i}}{h_{k}}}$,  for all $i=1, \dots, N$,  $\delta \in\{-1,1\}^{d}$   }
\vspace{0.5mm}
Calculate $\phi^j$ , $j=1,\dots, d$ such that $ x_{j,\phi^j(1)} \le  x_{j,\phi^j(2)} \le \dots \le x_{j,\phi^j(N)} $\\
\vspace{0.5mm}
$F(x_i,\delta) =0 $ for $i=1, \dots , N$,  for all $\delta \in \{-1,1\}^{d}$ \\
\vspace{0.5mm}
$\mathbf{RecurSplitting}(x, \psi,  \phi,  F, N)$\\
\KwOut{$ F(x_i,\delta)$ for all $i \in [1,N]$ and all $\delta \in \{-1,1\}^{d}$}
\vspace{1mm}
\caption{\label{algo:divideMain}Calculate  for $1 \le l \le m \le d$, $p$, $q$ given
$$  F(x_j, \delta) = \sum_{i=1}^N x_{l,i}^{p} x_{m,i}^{q} e^{\sum_{k=1}^{d}\frac{\delta_{k}x_{k,i}}{h_{k}}} \mathbbm{1}\{\delta_{1}x_{1,i} < \delta_{1}x_{1,j},\ldots,\delta_{d}x_{d,i} < \delta_{d}x_{d,j}\} , \quad j=1,N , \delta \in \{-1,1\}^{d} $$}
\end{algorithm2e}

\begin{algorithm2e}[H]
\DontPrintSemicolon 
\SetAlgoLined 
\vspace{1mm}
\KwIn{ $x$, $\psi$, $F$,   $\phi^j(i)$ for $i=1, M$, $j=1,d$}
\vspace{0.5mm}
$\kappa_1 = \{ \phi^{d}(i), i =1 , \frac{M}{2} \}$, $\phi_1$ with values in $\kappa_1$ s.t. $ x_{j,\phi_1^j(1)} \le  x_{j,\phi_1^j(2)} \le \dots \le x_{j,\phi_1^j(\frac{M}{2})}$, $j=1,d$\\
\vspace{0.5mm}
$\kappa_2 = \{ \phi^{d}(i), i = \frac{M}{2}+1 , M \}$, $\phi_2$  in $\kappa_2$ s.t. $ x_{j,\phi_2^j(1)} \le  x_{j,\phi_2^j(2)} \le \dots \le x_{j,\phi_2^j(\frac{M}{2})}$,$j=1,d$ \\
\vspace{0.5mm}
$\mathbf{RecurSplitting}(x, \psi,  \phi_1,  F, M/2)$\\
\vspace{0.5mm}
$\mathbf{RecurSplitting}(x, \psi,  \phi_2,  F, M/2)$ \\
\eIf{$\textcolor{blue}{\mathbf{(}}$ d > 2 $\textcolor{blue}{\mathbf{)}}$}
{
$\Delta = \{ \delta \in \{-1,1\}^{d} \}$\\
\vspace{0.5mm}
$\mathbf{MergeND}( x, \phi_1, \phi_2 , d-1, \psi,F, M/2, M/2 , \Delta)$
}
{
\Comment*[l]{ Merge for all $\delta$}
$\mathbf{Merge1D1}(x, \phi_1^1, \phi_2^1, \psi, F, \hat \Delta,M/2,M/2)$, with  $\hat \Delta = \{(1,1)\}$\\
\vspace{0.5mm}
$\mathbf{Merge1D2}(x, \phi_1^1, \phi_2^1, \psi, F, \hat \Delta,M/2,M/2)$  with  $\hat \Delta = \{(-1,1)\}$\\
\vspace{0.5mm}
$\mathbf{Merge1D1}(x, \phi_2^1, \phi_1^1, \psi, F,\hat \Delta,M/2,M/2) $  with  $\hat \Delta = \{(1,-1)\}$\\
\vspace{0.5mm}
$\mathbf{Merge1D2}(x, \phi_2^1, \phi_1^1, \psi, F,\hat \Delta,M/2,M/2)$  with  $\hat \Delta = \{(-1,-1)\}$\\
}
\KwOut{ $F$ updated }
\caption{\label{algo:divideAndConquer} Recursive splitting function $\mathbf{RecurSplitting}$}
\end{algorithm2e}
\clearpage
\begin{algorithm2e}[H]
\DontPrintSemicolon 
\SetAlgoLined 
\vspace{1mm}
\KwIn{ $x$, $\psi$, $F$, $\Delta \subset \{ \delta \in \{-1,1\}^{d}\} $ , $\phi_1^j(i)$, for all $i=1,M_1$, $\phi_2^j(i)$, for all $i=1,M_2$ with values in $[1,N]$ for $j=1,I_{dim}$}
\vspace{0.5mm}
\Comment*[l]{ $\delta \in \Delta$ s.t. $\delta_k x_k \le \delta_k y_k$ for $ k > I_{dim}$ for all $x \in \kappa_1$ and $y \in \kappa_2$ or   $\delta_k x_k >  \delta_k y_k$ for $ k > I_{dim}$ for all $x \in \kappa_1$ and $y \in \kappa_2$}
$\kappa_1 = \{ \phi_1^{I_{dim}}(i), i = 1 , M_1 \}$ , $\kappa_2= 
\{ \phi_2^{I_{dim}}(i), i = 1 , M_2 \}$\\
\vspace{0.5mm}
$\kappa =  \kappa_1 \cup  \kappa_2$, $x_{med}$ s.t; $\#\{ x_{j}, j \in \kappa ,  x_{I_{dim},j} \le x_{med}\} = \#\{ x_{j}, j \in \kappa ,  x_{I_{dim},j} > x_{med}\}$\\
\vspace{0.5mm}
$\kappa_{l,1} = \{ i \in \kappa_l , x_{I_{dim},i} \le x_{med}\}$,  $M_{l,1}= \# \kappa_{l,1}$, for $l=1,2$, \\
\vspace{0.5mm}
$\kappa_{l,2} = \{ i \in \kappa_l , x_{I_{dim},i} > x_{med}\}$,  $M_{l,2}= \# \kappa_{l,2}$,  for $l=1,2$ \\
Create $\phi_{l,m}^j(i)$, $i=1,\dots M_{l,m}$  s.t. $\phi_{l,m}^j(i) \in \kappa_{l,m}$, and $$x_{j,\phi_{l,m}^j(1)} \le  x_{j,\phi_{l,m}^j(2)} \le \dots \le x_{j,\phi_{l,m}^j(M_{l,m})}, \mbox{ for } j \le I_{dim}, \quad  l=1,2, \quad m=1,2 .$$\\
\vspace{0.5mm}
$\mathbf{MergeND}(x, \phi_{1,l}, \phi_{2,l} ,I_{dim}, \psi,F,  M_{1,l}, M_{2,l}, \Delta)$, for $l=1,2$\\
\vspace{0.5mm}
\eIf{$\textcolor{blue}{\mathbf{(}}$ $I_{dim}==2$ $\textcolor{blue}{\mathbf{)}}$}
{
\Comment*[l]{ Merge the set of 3D problem directly without recursion} 
$\mathbf{Merge1D1}(x, \phi_{1,1}^1, \phi_{2,2}^1, \psi, F, \hat \Delta, M_{1,1},M_{2,2})$ so that $\hat \Delta= \{ (1,1,1,..) \in  \Delta \}$\\
\vspace{0.5mm}
$\mathbf{Merge1D2}(x, \phi_{1,1}^1, \phi_{2,2}^1, \psi, F, \hat \Delta,M_{1,1},M_{2,2}) $ so that $\hat \Delta= \{ (-1,1,1,\dots) \in \Delta \}$\\
\vspace{0.5mm}
$\mathbf{Merge1D1}(x, \phi_{2,2}^1, \phi_{1,1}^1, \psi, F,\hat \Delta,M_{2,2},M_{1,1}) $ so that $\hat \Delta= \{(1,-1,-1,\dots) \in \Delta \}$\\
\vspace{0.5mm}
$\mathbf{Merge1D2}(x, \phi_{2,2}^1, \phi_{1,1}^1, \psi, F,\hat \Delta,M_{2,2},M_{1,1})$ so that $\hat \Delta= \{(-1,-1,-1,\dots) \in \Delta \}$\\
\vspace{0.5mm}
$\mathbf{Merge1D1}(x, \phi_{1,2}^1, \phi_{2,1}^1, \psi, F,\hat \Delta,M_{1,2},M_{2,1}) $ so that $\hat \Delta= \{(1,-1,1,\dots)  \in \Delta \}$\\
\vspace{0.5mm}
$\mathbf{Merge1D2}(x, \phi_{1,2}^1, \phi_{2,1}^1, \psi, F,\hat \Delta,M_{1,2},M_{2,1}) $ so that $\hat \Delta= \{(-1,-1,1,\dots) \in \Delta \}$\\
\vspace{0.5mm}
$\mathbf{Merge1D1}(x, \phi_{2,1}^1, \phi_{1,2}^1, \psi, F,\hat \Delta,M_{2,1},M_{1,2}) $ so that $\hat \Delta= \{(1,1,-1,\dots) \in \Delta \}$\\
\vspace{0.5mm}
$\mathbf{Merge1D2}(x, \phi_{2,1}^1, \phi_{1,2}^1, \psi, F,\hat \Delta,M_{2,1},M_{1,2}) $ so that $\hat \Delta= \{-1,1,-1,\dots\} \in \Delta \}$\\
}
{
\Comment*[l]{Merge in dimension below}
$\mathbf{mergedND}(x, \phi_{1,1}, \phi_{2,2} ,I_{dim}-1, \psi,F,  M_{1,l}, M_{2,l}, \hat \Delta)$, $\hat \Delta = \{ \delta \in  \Delta \mbox{ with }, \delta_{I_{dim}} \delta_{I_{dim}+1} >0 \}$\\
\vspace{0.5mm}
$\mathbf{mergedND}(x, \phi_{2,1}, \phi_{1,2} ,I_{dim}-1, \psi,F,  M_{1,l}, M_{2,l}, \hat \Delta)$, $\hat \Delta = \{ \delta \in  \Delta \mbox{ with }, \delta_{I_{dim}} \delta_{I_{dim}+1} <0\} $
}
\vspace{0.5mm}
\KwOut{ $F$ updated }
\caption{\label{algo:MergeND}Recursive merge nD  $\mathbf{MergeND}$ in given dimension $I_{dim}$}
\end{algorithm2e}
\clearpage
\begin{algorithm2e}[H]
\DontPrintSemicolon 
\SetAlgoLined 
\vspace{1mm}
\KwIn{ $x$, $\psi$, $F$,  $\phi_k$ s.t. $\phi_k(i) \le \phi_k(i+1)$, for all $i=1,M_k-1$, $k=1,2$, $\Delta  \subset \{ \delta \in \{-1,1\}^{d}\}$ }
$S(\delta) =0$ for all $\delta \in \Delta$,  $j=0$\\
\vspace{0.5mm}
\For{$\textcolor{blue}{\mathbf{(}}$ $i=1,M_2$ $\textcolor{blue}{\mathbf{)}}$}
{
\While{$\textcolor{blue}{\mathbf{(}}$ $(x_{\phi_2(i),1} \ge x_{\phi_1(j),1})$  and $j \le M_1$ $\textcolor{blue}{\mathbf{)}}$}
{
$S(\delta) +=  \psi( \phi_1(j),\delta)$ for all $\delta \in \Delta$ , $j= j+1$
}
$F(\phi_2(i) , \delta) += S(\delta)$ for all $\delta \in \Delta$ \\
\If{$\textcolor{blue}{\mathbf{(}}$ $j== M_1+1$ $\textcolor{blue}{\mathbf{)}}$}
{
\For{$\textcolor{blue}{\mathbf{(}}$ $k=i+1, M_2$ $\textcolor{blue}{\mathbf{)}}$}
{
$F(\phi_2(k) , \delta) += S(\delta)$ for all $\delta \in \Delta$ 
}
$i=M_2+1$
}
}
\KwOut{ $F$ updated }
\vspace{0.5mm}
\caption{\label{algo:Merge1D1} Final merge function in  dimension one : $\mathbf{Merge1D1}$ for two sets of points
$\kappa_1=\{ x_{\phi_1(i)}, i=1,M_1 \}$, $\kappa_2=\{ x_{\phi_2(i)}, i=1,M_2 \}$ such that for $x \in \kappa_1$, $y \in \kappa_2$,  $\delta_k x_{k} \le \delta_k y_{k} $ for $k \in [2,d]$, for all $\delta \in \Delta$. All elements $\delta$ of $\Delta$ are such that $\delta_1=1$. }
\end{algorithm2e}

\begin{algorithm2e}[H]
\DontPrintSemicolon 
\SetAlgoLined 
\vspace{1mm}
\KwIn{ $x$, $\psi$, $F$,  $\phi_k$ s.t. $\phi_k(i) \le \phi_k(i+1)$, for all $i=1,M_k-1$, $k=1,2$, $\Delta  \subset \{ \delta \in \{-1,1\}^{d}\}$ }
$S(\delta) =0$ for all $\delta \in \Delta$,  $j=M_1$\\
\vspace{0.5mm}
\For{$\textcolor{blue}{\mathbf{(}}$ $i=M_2,1$ $\textcolor{blue}{\mathbf{)}}$}
{
\While{$\textcolor{blue}{\mathbf{(}}$ $(x_{\phi_2(i),1} < x_{\phi_1(j),1})$  and $j \ge 1$ $\textcolor{blue}{\mathbf{)}}$}
{
$S(\delta) +=  \psi( \phi_1(j),\delta)$ for all $\delta \in \Delta$ , $j= j-1$
}
$F(\phi_2(i) , \delta) += S(\delta)$ for all $\delta \in \Delta$ \\
\If{$\textcolor{blue}{\mathbf{(}}$ $j==0$ $\textcolor{blue}{\mathbf{)}}$}
{
\For{$\textcolor{blue}{\mathbf{(}}$ $k=1, i-1$ $\textcolor{blue}{\mathbf{)}}$}
{
$F(\phi_2(k) , \delta) += S(\delta)$ for all $\delta \in \Delta$ 
}
$i=0$
}
}
\KwOut{ $F$ updated }
\vspace{0.5mm}
\caption{\label{algo:Merge1D2} Final merge function in dimension one : $\mathbf{Merge1D2}$ for two sets of points
$\kappa_1=\{ x_{\phi_1(i)}, i=1,M_1 \}$, $\kappa_2=\{ x_{\phi_2(i)}, i=1,M_2 \}$ such that for $x \in \kappa_1$, $y \in \kappa_2$,  $\delta_k x_{k} \le \delta_k y_{k} $ for $k \in [2,d]$, for all $\delta \in \Delta$. All elements $\delta$ of $\Delta$ are such that $\delta_1=-1$.}
\end{algorithm2e}

\end{document}